%% file: paper.tex
\PassOptionsToPackage{dvipsnames}{xcolor}

\documentclass[acmsmall,nonacm]{acmart}

\AtBeginDocument{%
  }

\usepackage{tikz}
\usepackage{placeins}
\usepackage{amsmath}
\newcommand{\papertablestyle}{%
  \centering
  \normalfont\fontsize{8}{9.5}\selectfont
  \setlength{\tabcolsep}{3.5pt}%
  \renewcommand{\arraystretch}{1.2}%
}
\usetikzlibrary{arrows.meta,positioning,shapes.geometric}
\makeatletter
\g@addto@macro\@floatboxreset{%
  \def\figure@type{figure}%
  \ifx\@captype\figure@type\sffamily\fi}
\makeatother
\makeatletter
\newcommand*{\da@rightarrow}{\mathchar"0\hexnumber@\symAMSa 4B }
\newcommand*{\xdashrightarrow}[2][]{%
  \mathrel{%
    \mathpalette{\da@xarrow{#1}{#2}{}\da@rightarrow{\,}{}}{}%
  }%
}
\newcommand*{\da@xarrow}[7]{%
  \sbox0{$\ifx#7\scriptstyle\scriptscriptstyle\else\scriptstyle\fi#5#1#6\m@th$}%
  \sbox2{$\ifx#7\scriptstyle\scriptscriptstyle\else\scriptstyle\fi#5#2#6\m@th$}%
  \sbox4{$#7\dabar@\m@th$}%
  \dimen@=\wd0
  \ifdim\wd2>\dimen@ \dimen@=\wd2\fi
  \count@=2
  \def\da@bars{\dabar@\dabar@}%
  \@whiledim\count@\wd4<\dimen@\do{%
    \advance\count@\@ne
    \expandafter\def\expandafter\da@bars\expandafter{%
      \da@bars\dabar@%
    }%
  }%
  #3%
  \mathop{\da@bars}\limits
  \ifx\\#1\\\else _{\copy0}\fi
  \ifx\\#2\\\else ^{\copy2}\fi
  #4%
}
\makeatother

\definecolor{pasoBlue}{RGB}{46,134,193}
\definecolor{pasoGreen}{RGB}{39,174,96}
\definecolor{pasoRed}{RGB}{231,76,60}
\definecolor{pasoOrange}{RGB}{243,156,18}
\definecolor{pasoGray}{RGB}{127,140,141}
\definecolor{pasoPurple}{RGB}{142,68,173}
\definecolor{codeBg}{RGB}{245,246,250}
\definecolor{keywordBlue}{RGB}{26,82,118}

\colorlet{cb-blue}{RoyalBlue}
\colorlet{cb-green}{ForestGreen}
\colorlet{cb-pink}{CarnationPink}
\colorlet{cb-red}{RedOrange}
\colorlet{cb-grey}{gray}
\colorlet{cb-cyan}{Cyan}

\def\lang{\mbox{\textsc{ProtoLang}}}
\newcommand{\meta}[1][automaton]{configuration-#1}

\usepackage{listings, listings-rust}
\lstdefinelanguage{paso}{
  keywords={while, if, else, for, prot, fn, struct, step, fork,
  yield, assert\_eq, X, fail, u1, u2, u8, u32, repeat, iterations,
  iters, uint, :=, is_last, iter_count, in},
  keywordstyle=\color{violet}\bfseries,
  otherkeywords={->,@},
  sensitive=true,
  comment=[l]{//}
}

\lstdefinelanguage{calyx}{
  keywords={invoke, else, static, import, if, while, par, seq, prim, wires, cells, group, with, control, component, ->, go, done, ?, primitive, verilog, external, repeat},
  keywordstyle=\color{darkgray}\bfseries,
  alsoletter=/->,
  sensitive=false,
  comment=[l]{//},
}
\definecolor{overviewFigureRed}{RGB}{181,23,0}    
\definecolor{overviewFigureBlue}{RGB}{0,77,128}  

\def\code{\lstinline[language=paso,basicstyle=\ttfamily]}

\title{Compiling Imperative Hardware Protocol Specifications\\ to Hardware}

\author{Nikil V. Shyamsunder}
\affiliation{
  \institution{Cornell University}
  \city{Ithaca}
  \state{NY}
  \country{USA}
}
\email{nvs26@cornell.edu}

\author{Ernest Ng}
\affiliation{
  \institution{Cornell University}
  \city{Ithaca}
  \state{NY}
  \country{USA}
}
\email{eyn5@cornell.edu}

\author{Adrian Sampson}
\affiliation{
  \institution{Cornell University}
  \city{Ithaca}
  \state{NY}
  \country{USA}
}
\email{asampson@cs.cornell.edu}

\author{Kevin Laeufer}
\affiliation{
  \institution{Cornell University}
  \city{Ithaca}
  \state{NY}
  \country{USA}
}
\email{laeufer@cornell.edu}

\begin{abstract}

Correct modular composition in hardware depends on interfaces with subtle multi-cycle timing behavior, yet such protocols remain difficult to specify, test, and debug.
Recent work has demonstrated that hardware communication specifications can take the form of compact imperative programs that describe the cycle-by-cycle behavior of each logical transaction.
However, prior tools only use these specifications for simulation-based testing and debugging,
and their semantics remain informal.
In this work, we develop a theory of hardware protocol specifications using automata.
In addition to making the specifications' meaning precise,
the automaton representation lets us compile them into synthesizable hardware.
We implement a compiler that produces RTL modules for \emph{driving} hardware interfaces and for \emph{reconstructing} logical transactions from low-level I/O signals.
We demonstrate a variety of use cases for synthesizable protocol drivers and reconstructors,
including runtime monitors that check security properties in the Wishbone protocol
and a binding generator that synthesizes interfaces between RTL modules and programs in the Calyx intermediate language.
\end{abstract}

\providecommand{\cupdot}{\mathbin{\mathpalette\pasocupdot{}}}
\newcommand{\pasocupdot}[2]{%
  \ooalign{$#1\cup$\cr\hidewidth$#1\cdot$\hidewidth\cr}%
}
\providecommand{\bigcupdot}{\mathop{\mathpalette\pasobigcupdot{}}\displaylimits}
\newcommand{\pasobigcupdot}[2]{%
  \ooalign{$#1\bigcup$\cr\hidewidth$#1\cdot$\hidewidth\cr}%
}

\begin{document}
\setlength{\textfloatsep}{12pt plus 2pt minus 2pt}
\maketitle

\makeatletter
\fancyhead[LO]{\ACM@linecountL}%
\fancyhead[RE]{\ACM@linecountR}%
\makeatother

\section{Introduction}

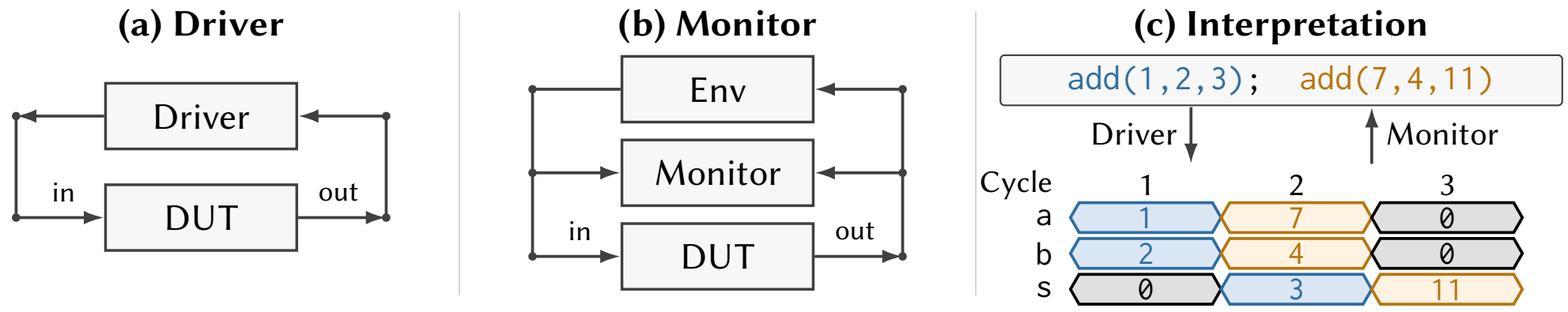
\begin{figure}[!b]
\centering
\input{fig/driver-monitor}
\caption{A driver (a) drives inputs and observes outputs on the design under test (DUT); a monitor (b) observes both.
Panel (c) shows their complementary mappings between transactions and signal traces.}
\label{fig:driver-monitor}
\end{figure}

Software functions expose an interface that largely hides how their implementations execute.
A caller supplies arguments and receives a result following a prescribed calling convention.
Hardware interfaces are more complex because of the centrality of cycle-level timing concerns.
A caller must drive and sample signals at particular clock cycles, observe handshakes, and respect limits on overlapping operations.
Hardware interfaces are inextricably linked with efficiency and functionality constraints:
for example, they reflect the depth and initiation interval of pipelined logic.
Hardware interfaces are therefore far more diverse than software calling conventions.

This variability creates an abstraction gap between \emph{transactions}, logical units of communication akin to function calls, and cycle-by-cycle behavior at the level of physical signals.
A given \emph{hardware communication protocol} describes how to implement transactions by varying signals over time.

Tragically, it is common to specify hardware protocols in natural language~\cite{Wishbone:2010, ARM:AXI:2024, ARM:AMBA:1999, arm_axi_stream_2021}.
Modular hardware engineering therefore requires \emph{ad hoc} techniques to check protocol conformance.
For example, engineers often separately develop
simulation testbenches that implement transactions for testing,
trace analyzers that recover transaction events from low-level signals,
temporal-logic descriptions for formal verifiers,
and hardware checkers to detect protocol violations in deployment.
Figure~\ref{fig:driver-monitor} illustrates two essential formulations for these techniques:
a \emph{driver} translates transactions into cycle-level behavior over time,
and a \emph{monitor} recognizes these cycle-level implementations to reconstruct corresponding transactions.

\begin{figure*}[t]
  \centering
  \includegraphics[width=\textwidth]{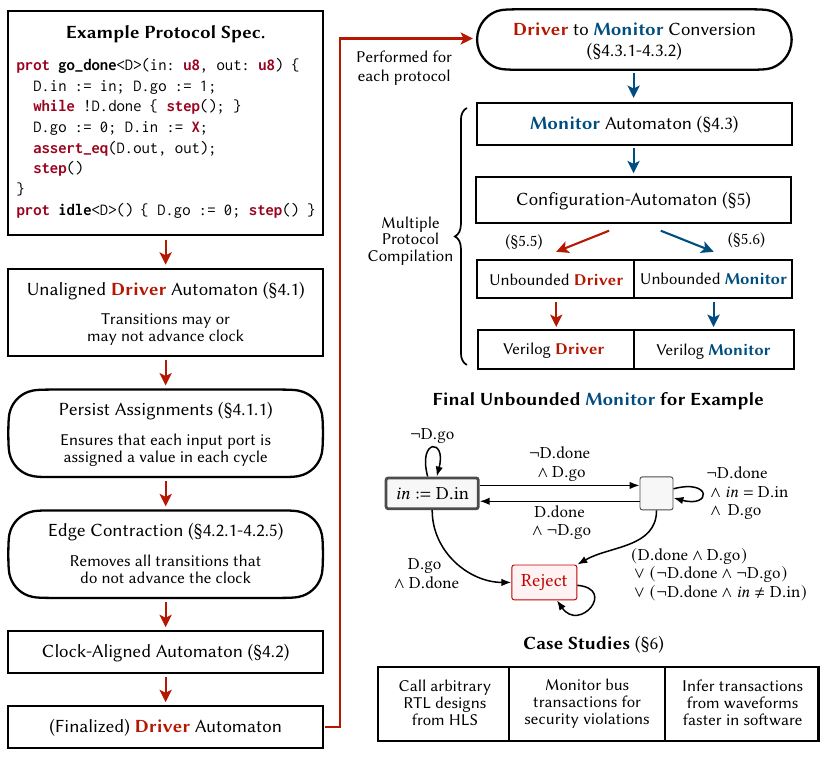}
  \caption{
    Our compiler from protocol specifications to automata-based drivers and monitors, on an example Go-Done protocol inspired by Calyx \cite{nigam2021calyx}.
    \textcolor{overviewFigureRed}{Red} and \textcolor{overviewFigureBlue}{blue} arrows 
    indicate \textcolor{overviewFigureRed}{driver} and \textcolor{overviewFigureBlue}{monitor}-relevant passes respectively. Rectangles denote intermediate representations; rounded boxes represent compiler passes.
  }
  \label{fig:overview}
\end{figure*}

Recent work has aimed to unify these disparate tools using machine-readable protocol specifications~\cite{ng_et_al_2026}.
This strategy uses a domain-specific language (DSL) to describe a hardware protocol as a compact imperative program that resembles a driver.
That prior work has demonstrated how to use the same specification for trace analysis, so a single source of truth provides both a driver and a corresponding monitor. However, it has two main limitations.
First, the protocol specification DSL has only informal semantics.
Second, the driver and monitor tools target only software simulation of hardware designs.
There is no way to translate a protocol specification into a hardware implementation,
which would be useful for implementing correct-by-construction hardware interfaces and for in-deployment conformance monitoring.

We address both problems by formalizing hardware protocol specifications as finite automata.
We introduce an automaton representation and show how to translate specifications written in the \lang{} DSL~\cite{ng_et_al_2026} into these automata.
We then implement a compiler that translates protocol automata into simple finite state machines (FSMs) that admit direct translation into register-transfer level (RTL) hardware descriptions.
Our compiler can produce both \emph{driver} implementations (Figure~\ref{fig:driver-monitor}(a)) and \emph{monitor} implementations (Figure~\ref{fig:driver-monitor}(b)) from a single \lang{} source program.

Compiling imperative specifications into hardware presents two main challenges.
First, drivers and monitors must handle unbounded transaction traces, but we need to implement them using fixed, finite storage resources.
Second, we need to translate the imperative specification's sequential statements into parallel hardware behavior while preserving their semantics.
We design a series of compiler passes that lower \lang{} protocols into automata representing a single transaction,
pack their logic into cycle-aligned steps,
translate them from drivers to monitors,
and compose the resulting automata to handle unbounded transaction sequences.

Our key contributions are:
\begin{itemize}
  \item We introduce \emph{protocol automata}, a formal foundation for describing hardware communication protocols. We show how to translate the \lang{} DSL~\cite{ng_et_al_2026} into protocol automata.
  \item We statically transform protocol automata into hardware-implementable automata that act as either \emph{drivers}, which implement transaction behavior, or \emph{monitors}, which reconstruct transaction parameters and check protocol compliance.
  \item We characterize protocols whose monitoring and concurrent execution admit finite-state automata over infinite traces, and give a terminating construction.
  \item We apply our compiler in three use cases:
  synthesizing low-overhead in-hardware monitor logic that checks security violations;
  generating bindings between RTL modules and a high-level synthesis (HLS) compiler toolchain;
  and high-performance trace analysis.
\end{itemize}

\section{Compilation by Example}
\label{sec:motivating-example}
Consider an ALU that supports 32-bit addition and multiplication.
It has two operand ports, \code|a| and \code|b|, and a result port, \code|s|.
A 1-bit opcode input \code|op| selects between addition (0) and multiplication (1).
The adder has a pipeline of depth 1 and therefore a fixed-latency protocol: it returns a result one cycle after receiving its operands and can accept new operands every cycle.

In contrast, suppose the multiplier is iterative and can terminate early for small inputs, giving it data-dependent latency.
This variable latency requires the multiplier to signal completion explicitly to its caller.
We choose an interface using two 1-bit signals, \code|go| and \code|done|, inspired by Calyx~\cite{nigam2021calyx}:
\begin{itemize}
  \item The caller starts an operation by supplying its inputs and asserting \code|go|.
  \item While \code|done| is low, the caller holds \code|go| high and keeps the operation's inputs stable.
  \item The design raises \code|done| to indicate that the operation has completed.
  \item In the same cycle, the caller samples the outputs and lowers \code|go|. The inputs are released, and a new operation may start in the following cycle.
\end{itemize}

\paragraph{Protocols as programs.}
Figure~\ref{fig:alu-source} specifies the \texttt{add} and \texttt{mul} protocols in \lang{}~\cite{ng_et_al_2026}.~\footnote{We refer the reader to \citet{ng_et_al_2026} for the full syntax of \lang{}}
The parameters \code|a| and \code|b| represent the operands we wish to add or multiply, and \code|s| is the expected result.
The handle \code|D| provides access to the ports of the design under test (DUT): \code|D.p| refers to port \code|p|.
An assignment \code|D.a := a| drives port \code|a| with the value of the right-hand side expression.
After driving the inputs, the \code|add| transaction must wait one cycle to read the output; \code|step| advances one clock cycle.
The assertion \code|assert_eq(D.s, s)| checks the computed result matches the transaction parameter.
Note that protocols do not specify \emph{functional} correctness, i.e., they do not check that \code|s| is the sum or product of \code|a| and \code|b|.
The assertions only specify how to execute protocol transactions.

The \code|fork| statement allows another transaction to start, overlapping with the current transaction.
In \code|add|, another operation may begin in the second cycle.
\lang{} assignments persist until they are overwritten,
so protocols need to explicitly end assignments to let other transactions use the DUT's ports.
Using the special symbol \code|X| on an assignment's right-hand side releases the port.

Unlike \code|add|, \code|mul|'s data-dependent latency cannot be expressed by a fixed number of \code|step| calls.
It therefore uses a \code|while| loop to wait for \code|D.done| to become high.
While waiting, assignment persistence means that the operands, opcode, and \code|go| remain stable.
Once \code|D.done| rises, the protocol lowers \code|D.go|, releases the operands and opcode, and checks the result.
When a transaction has no explicit \code|fork|, the next transaction can start after the current one finishes.
\input{fig/alu-source}

\paragraph{From source programs to automata.}
We represent \lang{} programs as \emph{protocol automata}.
The states contain assignments, assertions, and forks (\S\ref{sec:automata-actions}); edges denote control flow and the passage of time.
Figure~\ref{fig:alu-unaligned} shows \code|mul| as an automaton.
The \emph{actions} within a state execute on transitioning out of that state.
Each edge is either a \emph{step edge}, which advances the DUT clock by one clock cycle, or a \emph{nonstep edge}, which does not advance time (\S\ref{sec:small-step}).

Because \lang{} specifications are imperative programs,
translating them to protocol automata is straightforward and resembles traditional control-flow graph (CFG) construction.
Nonstep edges are a crucial factor in representing sequential composition within a single clock cycle.

Whenever no outgoing transition guard holds, execution takes an implicit transition to a self-looping rejecting state. 
In the automata constructed here, each exit state has no outgoing transition, so continuing after the transaction completes enters this rejecting state. Thus, these automata represent a \emph{single transaction}, rejecting extra cycles beyond those specified by the protocol.

\input{fig/alu-unaligned}

\paragraph{Removing nonstep edges.}
Our compiler's goal is to construct automata that transition exactly once per clock cycle, enabling efficient hardware implementation.
The first step is therefore to remove nonstep edges, producing a \emph{clock-aligned} automaton (\S\ref{sec:edge-contraction}).
The clock-aligned automaton will need to have multiple actions per state.
To match hardware semantics, the actions within a state run in parallel---but we also need to preserve the original automaton's sequential effects.

Figures~\ref{fig:alu-automata}(a) and~(b) illustrate clock-aligned automata for \code|add| and \code|mul|.
Here, assignments are conditional:
$v := \{(g_1, e_1), (g_2, e_2), \dots\}$
assigns $v$ to the value of $e_i$ if the predicate $g_i$ holds (\S\ref{sec:expression-semantics}).
In \code|mul|'s waiting state (1),
the assignments depend on
\code|D.done|: when false, the inputs remain driven; when true, the operands are released and \code|D.go| is lowered.
Similarly, $[g]\;a$ denotes an action that executes only when $g$ is true (\S\ref{sec:action-conditioning}); thus, the result assertion applies only when \code|D.done| holds.

\input{fig/alu-individual-automata}

\paragraph{Reconstructing transactions.}
The automata so far act as \emph{drivers};
we also want to produce \emph{monitors}.
The goal is an automaton that uses assignments and assertions to learn values for transaction parameters (\S\ref{sec:monitor-automata}).
We construct monitor automata that assign to \emph{register} variables corresponding to each parameter.
Figures~\ref{fig:alu-automata}(c) and~(d) show the monitors for \code|add| and \code|mul|.
Registers $v_a$, $v_b$, and $v_s$ contain the learned values of $a$, $b$, and $s$.
In \code|mul|, the first cycle checks that \code|op| is 1 and that \code|go| is high, and it learns the operands.
While waiting, the monitor checks that the operands and control inputs remain stable; on completion, it checks that \code|go| is low and learns the result in $v_s$.

\paragraph{Unbounded transaction streams.}
Our automata above each recognize a single transaction.
We need automata that handle unbounded executions
with arbitrary interleavings of \code|add| and \code|mul|.
We compose the single-transaction monitors by exploring which transactions can execute together and when new ones may begin.
We call the result a \emph{\meta{}} (\S\ref{sec:meta-automata}).
By recognizing recurring configurations and reusing storage from retired transactions, we obtain a finite representation of unbounded traces (\S\ref{sec:steady-states}).
We populate this representation with the individual monitors' actions and constraints to produce one hardware monitor for the combined protocol stream (\S\ref{sec:unbounded-monitors}).
Figure~\ref{fig:alu-unbounded-monitor} illustrates the result for \code|add| and \code|mul|.
\input{fig/alu-unbounded-monitor}

\section{Protocol Automata}
\label{sec:paso-automata}
This section formalizes the protocol automaton representation that we use throughout compilation.

\begin{definition}[Protocol automaton]
\label{def:paso-automaton}
A \emph{protocol automaton} is a tuple
\[
  \mathcal A=(Q,q_{\mathrm{entry}},R,E,\mathcal{I},\delta).
\]
Here $Q$ is a finite set of states, each carrying a finite set of actions, and $q_{\mathrm{entry}}\in Q$ is the entry state.
The disjoint finite symbol sets $R$ and $E$ contain internal registers and external signals, respectively, each with a type (i.e., a bitwidth).
We refer to the set of all symbols $S=R\cup E$.
(These sets are fixed;
the automaton cannot dynamically allocate storage.)
The $\mathcal{I}$ function assigns each register in $R$ an initial bitvector value. 
The transition function $\delta$ assigns each state a finite set of outgoing transitions $(g,k,q')$, where $g$ is a Boolean predicate, $k\in\{\mathrm{nonstep},\mathrm{step}\}$, and $q'\in Q$ is the target.

All states in $Q$ are accepting states.
The automaton rejects whenever no outgoing transition guard holds.
This implicit rejection makes the transition relation total, so $\mathcal A$ is deterministic when every state's outgoing guards are pairwise disjoint.
Compilation will maintain this invariant.
\end{definition}

\subsection{Expressions}
\label{sec:expression-semantics}
Expressions $e$ in our protocol automata include symbol references drawn from $S$, and the usual Boolean and bitvector operators including bitwise logical operators and shifts.
Let $\mathcal B_n$ be the set of $n$-bit vectors, and let $\mathcal B_n^X=\mathcal B_n\cup\{X\}$ additionally contain the distinguished ``don't care'' marker.
Let $\Sigma$ denote the set of environments: total functions from $S$ to $\mathcal B_n^X$ for the appropriate bitwidth.
An environment $\sigma\in\Sigma$ records the current values of registers and external signals.
An $n$-bit expression has a partial denotation $\llbracket e\rrbracket:\Sigma\rightharpoonup\mathcal B_n^X$.
Registers initially have the values given by $\mathcal{I}$, while the surrounding hardware supplies the initial external-signal values.
Ordinary expressions have their standard bitvector denotations.
For a symbol $s$, $\llbracket s\rrbracket(\sigma)=\sigma(s)$ only when $\sigma(s)\ne X$; reading a symbol whose value is $X$ is undefined.
The expression \code|X| denotes the marker itself: $\llbracket\texttt{X}\rrbracket(\sigma)=X$.
All operations over expressions are undefined if an operand is undefined or evaluates to $X$.

Figure~\ref{fig:alu-automata} illustrates how removing nonstep edges introduces assignments whose values depend on guards. Thus, we extend the expression syntax for automata with one extra constructor to represent these conditional expressions, called \emph{value summaries}.

\begin{definition}[Value summary]
\label{def:value-summary}
A \emph{value summary} is an expression
\[
  V=\{(g_1,e_1),\ldots,(g_k,e_k)\},
\]
where guards $g_i$ are Boolean expressions over $S$,
and expressions $e_i$ have the same bitwidth.
The guards must be pairwise disjoint.
The denotation is the value of the unique enabled expression:
\[
  \llbracket V\rrbracket(\sigma)
    =\llbracket e_i\rrbracket(\sigma)
    \quad\text{if }\llbracket g_i\rrbracket(\sigma)=1.
\]
A summary is \emph{total} when $\models\bigvee_{i=1}^{k}g_i$.
Its denotation is undefined if no guard holds or the selected expression is undefined.
Note that the assignment $s:=e$ is simply syntactic sugar for $s:=\{(1,e)\}$.
Value summaries are inspired by those used in symbolic execution~\cite{sen_et_al_2015}.
\end{definition}

\subsection{Actions}
\label{sec:automata-actions}

The action syntax is
\[
  a ::= s:=V
    \mid \text{\code|assert_eq|}(e_1,e_2)
    \mid \text{\code|fail|}
    \mid \text{\code|fork|},
  \qquad s\in S.
\]
Namely, actions correspond to atomic statements in \lang{}.
The compiler introduces \code|fail| to reject violations of \lang{} invariants (e.g., in Appendix~\ref{sec:fork-instrumentation}), whereas assertions originate in the source program.

Let $\mathcal U$ be the set of \emph{updates:}
partial functions from $S$ to $\mathcal B_n^X$ for some $n$.
Write $[\,]$ for the empty update.
Applying an update to an environment overwrites its target, preserving every other symbol:
\[
  (\sigma\oplus u)(s)=
  \begin{cases}
    u(s) & s\in\operatorname{dom}(u),\\
    \sigma(s) & \text{otherwise}.
  \end{cases}
\]
We give each action a partial denotation $\llbracket a\rrbracket:\Sigma\rightharpoonup\mathcal U$, mapping the incoming environment to an update.
An assignment records the value of its right-hand side:
\[
  \llbracket s:=V_s\rrbracket(\sigma)
    =[s\mapsto\llbracket V_s\rrbracket(\sigma)].
\]
An assertion produces the empty update if both operands denote the same concrete bitvector:
\[
  \llbracket\text{\code|assert_eq|}(e_1,e_2)\rrbracket(\sigma)
    =[\,]
    \quad\text{if }\llbracket e_1\rrbracket(\sigma)=\llbracket e_2\rrbracket(\sigma)\in\mathcal B_n.
\]
Its denotation is undefined otherwise.
The denotation of \code|fail| is undefined for every environment.

The \code|fork| action is a no-op in single-protocol execution, but remains a marker for later composing multiple transactions (\S\ref{sec:meta-automata}). Thus, fork always produces the empty update:
\[
  \llbracket\text{\code|fork|}\rrbracket(\sigma)=[\,].
\]

\paragraph{Guarded actions.}
A \emph{guarded action}, written $[g]a$, executes action $a$ exactly when the parameter-free guard $g$ holds.
We will occasionally use $a$ as syntactic sugar for $[1]a$.
The denotation of a guarded action is also a partial function from environments to updates:
\[
  \llbracket[g]a\rrbracket(\sigma)
    =\begin{cases}
      \llbracket a\rrbracket(\sigma) & \text{if }\llbracket g\rrbracket(\sigma)=1,\\
      [\,] & \text{if }\llbracket g\rrbracket(\sigma)=0.
    \end{cases}
\]

\paragraph{Evaluating multiple actions.}
Every automaton state corresponds to a set of actions, which execute in parallel.
Semantically, we evaluate every action within the same incoming environment $\sigma$ and then combine their updates to produce a new environment.
Write $u\cupdot v$ for the union of updates with disjoint domains (which is undefined when the domains overlap).
For a set of actions $A$, define
\[
  \llbracket A\rrbracket(\sigma)
    =\bigcupdot_{a\in A}\llbracket a\rrbracket(\sigma).
\]
For a state $q$ carrying actions $A$, applying the combined update gives its outgoing environment:
\[
  \llbracket q\rrbracket(\sigma)
    =\sigma\oplus\llbracket A\rrbracket(\sigma).
\]

\subsection{Small-step semantics}
\label{sec:small-step}

Protocol automata take two kinds of transitions:
nonstep transitions stay in the current clock cycle,
and step transitions advance the DUT clock before the target state executes.
We assume a model of the surrounding hardware $\mathcal H$ that updates the external signals $E$ for the next cycle, respecting port assignments and the stability of parameter signals while preserving the register values in $R$.
On a nonstep transition, the surrounding hardware supplies no new signal values.

We define a small-step operational semantics over pairs $(q,\sigma)$, where $q\in Q$ is the current automaton state and $\sigma\in\Sigma$ is the current environment:
\[
  \textsc{Nonstep}\enspace
  \frac{
    \begin{gathered}
      (g,\mathrm{nonstep},q')\in\delta(q)\quad\llbracket g\rrbracket(\sigma)=1\\
      \llbracket q\rrbracket(\sigma)=\sigma'
    \end{gathered}
  }{(q,\sigma)\xrightarrow{\mathcal A,\mathrm{nonstep}}(q',\sigma')}
  \qquad
  \textsc{Step}\enspace
  \frac{
    \begin{gathered}
      (g,\mathrm{step},q')\in\delta(q)\quad\llbracket g\rrbracket(\sigma)=1\\
      \llbracket q\rrbracket(\sigma)=\tau\quad(\tau,\sigma')\in\mathcal H
    \end{gathered}
  }{(q,\sigma)\xrightarrow{\mathcal A,\mathrm{step}}(q',\sigma')}
\]
Execution starts at $q_{\mathrm{entry}}$ with registers initialized by $\mathcal{I}$.
If no outgoing guard holds, or the source state's evaluation is undefined, execution enters the implicit rejecting state and remains there.

\section{Single-Protocol, Single-Transaction Compilation}
\label{sec:single-protocol-compilation}
We compile a protocol in two stages:
we first produce a general automaton
and then remove all nonstep edges to produce a \emph{clock-aligned} automaton.
\subsection{Lowering a \lang{} Source Protocol to an Automaton}
Figure~\ref{fig:ast-to-ir} illustrates a recursive translation from
a \lang{} \code|prot| definition's body to a naive automaton.
We add an entry state $q_{\mathrm{entry}}$, which
sets every DUT input port to \code|X|,
and an exit state $q_{\mathrm{exit}}$.
Reaching $q_{\mathrm{exit}}$ marks completion in the cycle \emph{after} the transaction's final live cycle.
The translation of \code|repeat| and \code|for-in| loops is similar to the \code|while| case;
Appendix~\ref{app:iterator-lowering} provides the details.

\input{fig/ast-to-ir}

\subsubsection{Persisting Assignments}
\label{sec:persist-assignments}

In \lang{}, a DUT input port retains its value until the protocol assigns it again.
To preserve this behavior in compiled hardware, we instrument the automaton with explicit registers and assignments.
This establishes the invariant that every input port is explicitly assigned in each cycle, simplifying the reasoning needed in subsequent compilation passes.

For each port \code|D.p|,
we introduce a value register $v_p$ and a release bit $x_p$, which is initially 1.
When $x_p=1$, the represented value is $X$;
when $x_p=0$, it is the value in $v_p$.
At a state that assigns a concrete expression \code|D.p := e|, add the parallel register updates $v_p:=e$, and $x_p:=0$.
At a state that assigns $p:=\texttt{X}$, set $x_p:=1$.
Figure~\ref{fig:ast-to-ir} shows that \code|step| translates to a step transition targeting an empty state.
We use this state to restore the previously assigned input values.
For every input port $p$, this state assigns
$
  p:=\{(x_p,\texttt{X}),(\neg x_p,v_p)\}.
$
Thus, each cycle starts with the saved input values, and later assignments in that cycle override them.
Figure~\ref{fig:adder-holdover} illustrates this instrumentation for \code|add|.
\input{fig/adder-holdover}

\subsection{Clock-Aligning Automata}
\label{sec:Clock-automata}
The next phase eliminates nonstep transitions, producing a \emph{clock-aligned} automaton.
It combines actions from the same cycle into a single state while
preserving their sequential effects.~\footnote{
The idea of using nonstep edges in the initial automaton construction and later contracting them was inspired by the SystemVerilog Assertion compiler (\texttt{verificsva.cc}) in the open-source Yosys tool~\cite{yosys_austrochip_2013}. Yosys calls nonstep edges \textit{links}.
}

\subsubsection{Edge Contraction}
\label{sec:edge-contraction}
We \emph{contract} nonstep edges forward from each \emph{cycle entry}.
A cycle entry is either the protocol entry or the state following a step transition. %
For each cycle entry $u$, process its nonstep successors in topological order, accumulating their effects in $u$.
To contract $u\xdashrightarrow{g}v$, merge its actions under the rules in \S\ref{sec:sequential-merge}.
For every outgoing transition $v\xrightarrow{g'}w$, add $u\xrightarrow{g\land g'}w$ with the same step or nonstep kind, then delete $u\xdashrightarrow{g}v$.
Continue until $u$ has only step transitions.
If a same-cycle path revisits a state, replace that cyclic continuation with a \code|fail| action guarded by its path condition.
This algorithm terminates because each explored path ends at a step transition, a terminal state, or a repeated state, so only finitely many path prefixes are explored.
We first develop the theory to correctly describe merging that preserves the sequential semantics of nonstep edges in a single, parallel execution state (\S\S\ref{sec:action-conditioning},\ref{sec:sequential-assignment-merge},\ref{sec:substitutions}), providing the complete merge rules in \S\ref{sec:sequential-merge}.

\subsubsection{Action Conditioning}
\label{sec:action-conditioning}
Contracting $u\xdashrightarrow{g}v$ combines two sequentially executed states into one whose actions execute in parallel.
The ordered merge must preserve the effects of executing $u$ before $v$, while activating $v$'s actions only when $g$ holds.
We express these conditional effects using \emph{guarded actions} and \emph{action conditioning}.
Define the operator $\cdot$ for conditioning an action on a guard:
\[
g\cdot[g']a \triangleq [g\land g']a.
\]
Assignments receive special treatment, where $g$ is pushed into the value summary:
\begin{gather*}
g\cdot V_p \triangleq \{(g\land g',e)\mid(g',e)\in V_p\}\\
g\cdot(p:=V_p) \triangleq p:=(g\cdot V_p)
\end{gather*}

Thus assignment actions retain an implicit guard of $1$, with all control flow represented \emph{within} their value summaries.
This special treatment is useful when performing substitutions in  \S\ref{sec:substitutions}.

\subsubsection{Sequential Assignment Merge}
\label{sec:sequential-assignment-merge}
When edge contraction combines assignments to the same symbol, their effect must reflect their original order.
Let $V_p$ be the value summary already in an earlier state, and let $V'_p$ come from a later state.
Define the coverage of $V'_p$ as
$
  c_{V'} \triangleq \bigvee_{(g,e)\in V'_p}g.
$
Their \emph{sequential assignment merge} is
\[
  V_p \mathbin{\blacktriangleright} V'_p
  \triangleq
  \operatorname{coalesce}\left(
    V'_p\cup\{(g\land\neg c_{V'},e)\mid(g,e)\in V_p\}
  \right).
\]
Here, $\operatorname{coalesce}(V)$ combines entries for the same expression by disjoining their guards.

\subsubsection{Substituting Earlier Assignments.}
\label{sec:substitutions}
To simulate the ordering enforced by nonstep edges in a single state, where actions execute in parallel, we must rewrite future uses of symbols with the assignments they are given by the previous state.
Figure~\ref{fig:wire-substitution} illustrates this for a combinational wire.
Let $V_p$ be a total value summary in the earlier state.
Before merging a later state, simultaneously substitute $p\mapsto V_p$ in all its expressions, for every symbol $p$ assigned in the earlier state.
Since assignment actions are always unconditional, this substitution is straightforward.

\input{fig/wire-substitution}

\subsubsection{Sequential Merge}
\label{sec:sequential-merge}
\input{fig/sequential-merge}
Table~\ref{tab:sequential-merge} lists the rules for sequential merge.
Each state contains at most one \code|fail|, one \code|fork|, and one value summary per assignable symbol.
We therefore match incoming \code|fail| and \code|fork| actions by kind, and assignments by their target symbol.
Merging an assignment replaces the earlier summary $V_p$ with the resulting ordered merge, which is used in subsequent substitutions and contractions.
Assertions are always inserted as separate actions.

\subsubsection{Automata Simplification}
\label{sec:simplify}
We simplify guards and value summaries using constant folding and algebraic rewrites, reducing combinational logic and sometimes detecting or ruling out failures statically.
A forward dataflow pass, akin to constant and copy propagation, replaces register reads with the expressions they were directly assigned.
For this to be sound across cycles, we only propagate expressions over constants and parameters, which are immutable across cycles, as opposed to references to DUT ports, which are mutable by the external hardware.
Finally, dead-code elimination removes unused symbols.
The resulting drivers are given in Figure~\ref{fig:alu-automata}.

\subsection{Converting Drivers to Monitors}
\label{sec:monitor-automata}

The compiler flow so far produces \emph{driver} automata;
we optionally transform these into \emph{monitor} automata.
This conversion has two phases:
we transform all assignments and assertions into constraints (\S\ref{sec:constraints});
and we insert registers that ``learn'' bits in transaction parameters
(\S\ref{sec:parameter-knowledge}).

\subsubsection{Constraints}
\label{sec:constraints}

A monitor automaton observes both DUT input and output ports.
Therefore,
assignments and assertions impose the same requirement: the observed values must agree with the protocol's expressions.
We therefore rewrite both kinds of action as \emph{constraints}.

A \emph{constraint}, written $[g]\;e_1=e_2$, requires expressions $e_1$ and $e_2$ of the same bit width to agree whenever the parameter-free guard $g$ holds.
Constraints are symmetric: information can flow left to right or right to left, depending on which side contains an unknown parameter.
We would like to normalize these constraints so that information flows only right to left, from a directly observable expression to a parameter on the left, or does not flow at all.
A \emph{parameter constraint} has the form $ [g]\;a[h:l]=\mathit{word}, $ where $a$ is an $n$-bit parameter, $0\le l\le h<n$, and the expression $\mathit{word}$ is parameter-free.
A \emph{parameter-free constraint} has the form $[g]\;e_1=e_2$, where $g$, $e_1$, and $e_2$ are \emph{all} parameter-free.
It checks consistency without learning new parameter values.

\paragraph{\lang{} Parameter Restrictions}
We consider three \lang{} restrictions that make normalizing into parameter or parameter-free constraints tractable. 
(1) Control flow cannot branch on transaction parameters, so guards can be evaluated directly without first learning parameter values. In other words, guards are always parameter-free.
(\code|for|--\code|in| loops are a separate case that we handle in Appendix~\ref{app:monitor-iterations}.)
(2) Each expression contains at most one parameter occurrence, giving normalization a single parameter-bearing operand to isolate.
(3) Assignments and assertions contain parameters on at most one side.
Assignments naturally produce constraints of this form, relating an observed DUT input to the expression driving it.
Because the caller chooses transaction arguments, assertions such as \code|assert_eq(p, q)| between parameters express \emph{functional} relationships rather than communication behavior, so it is natural to disallow them.

We structurally recurse through each constraint's expressions using the normalization rules in Table~\ref{tab:constraint-normalization} until every remaining constraint is either a parameter constraint or parameter-free.
The common guard $g$ is inherited unless a row gives a new guard.
An unsliced $n$-bit expression denotes its full range $[n-1:0]$.
Let $r$ be the width of $e_2$ in these rules.

\input{fig/isolate-params}

\subsubsection{Learning Parameters from Observations}
\label{sec:parameter-knowledge}

Each state's driver actions generate constraints in the current cycle, but all cycles must agree on the transaction's immutable parameter values.
We translate these constraints \emph{back} into automaton actions, using register assignments to learn parameter bits across cycles and assertions to check later observations agree with what is known.

For each $n$-bit parameter $a$, the monitor encodes its knowledge of $a$ in two $n$-bit registers, a value $v_a$ and a known-bit mask $k_a$.
Bit $k_a[i]=1$ means that $a[i]$ is known and equals $v_a[i]$; when $k_a[i]=0$, the corresponding value bit imposes no constraint.
Their initial values are both zero.

Within a state, we will accumulate a value summary to update $v_a$ and $k_a$, beginning with $\mathit{Value}=\{(1,v_a)\}$ and $\mathit{KnownBits}=\{(1,k_a)\}$ (the values from the previous cycle).
Define $\operatorname{zext}_n$ as a function that zero-extends a bitvector to width $n$, and $\operatorname{ones}(b)$ as a $b$-bit vector of ones.

For each parameter constraint $[g]\;a[h:l]=\mathit{word}$ in the state, update $\mathit{Value}$ and $\mathit{KnownBits}$.
First form the observation mask $\mathit{mask}_{\mathrm{obs}}$, which is 1 in bits $h$ to $l$, and align the observed word $\mathit{word}_{\mathrm{obs}}$ to the parameter's $n$-bit width:
\[
  \mathit{mask}_{\mathrm{obs}}\triangleq\operatorname{zext}_n(\operatorname{ones}(h-l+1))\ll l,\qquad
  \mathit{word}_{\mathrm{obs}}\triangleq\operatorname{zext}_n(\mathit{word})\ll l.
\]
Of the observed bits, $\mathit{mask}_{\mathrm{seen}}$ marks known bits and $\mathit{mask}_{\mathrm{new}}$ marks newly learned bits.
\[
  \mathit{mask}_{\mathrm{seen}}\triangleq \mathit{KnownBits}\mathbin{\&}\mathit{mask}_{\mathrm{obs}},\qquad
  \mathit{mask}_{\mathrm{new}}\triangleq \mathit{mask}_{\mathrm{obs}}\mathbin{\&}\neg \mathit{KnownBits}.
\]
For each observed bit that is already known, check that its newly observed value agrees with the value recorded in $\mathit{Value}$ by comparing the two words under the mask $\mathit{mask}_{\mathrm{seen}}$:
\[
  [g]\;\operatorname{assert\_eq}
    (\mathit{Value}\mathbin{\&}\mathit{mask}_{\mathrm{seen}},\;
     \mathit{word}_{\mathrm{obs}}\mathbin{\&}\mathit{mask}_{\mathrm{seen}}).
\]
Then extend the value summaries using ordered merge, combining bits of $\mathit{Value}$ outside $\mathit{mask}_{\mathrm{new}}$ with newly observed bits of $\mathit{word}_{\mathrm{obs}}$ inside it, marking all observed bits as known:
\[
\begin{aligned}
  \mathit{Value}'&\triangleq \mathit{Value}\mathbin{\blacktriangleright}
    \{(g,(\mathit{Value}\mathbin{\&}\neg \mathit{mask}_{\mathrm{new}})
      \mathbin{|}(\mathit{word}_{\mathrm{obs}}\mathbin{\&}\mathit{mask}_{\mathrm{new}}))\},\\
  \mathit{KnownBits}'&\triangleq \mathit{KnownBits}\mathbin{\blacktriangleright}
    \{(g,\mathit{KnownBits}\mathbin{|}\mathit{mask}_{\mathrm{obs}})\}.
\end{aligned}
\]
When $g$ is false, the updates preserve the preceding value and known-bit summaries.

Multiple constraints in one state must agree with one another and with values learned in earlier cycles.
We therefore process them in order while constructing the monitor.
The summaries $\mathit{Value}',\mathit{KnownBits}'$ produced by each constraint are the inputs to the next constraint, while its assertion uses the summaries from before its update.
This ordering builds expressions rather than performing intermediate register writes: after processing all $c$ constraints, we emit only $v_a:=\mathit{Value}^{(c)}$ and $k_a:=\mathit{KnownBits}^{(c)}$ using the final summaries. Automata simplification (\S\ref{sec:simplify}) can often simplify away these $v$ and $k$ registers.

Figure~\ref{fig:monitor-examples} illustrates the conversion for a driver state assigning overlapping slices of a five-bit parameter $p$ to two three-bit DUT inputs.
The assignments give the parameter constraints $p[3:1]=\mathrm{D.a}$ and $p[4:2]=\mathrm{D.b}$, both with guard $1$.
Initially $v_p=k_p=0$.
Processing the first constraint learns bits $p[3:1]$: $\mathit{mask}_{\mathrm{obs}}=\texttt{01110}_2$ and $\mathit{mask}_{\mathrm{seen}}=0$, so no consistency check is needed.
The accumulated value now contains $\mathrm{D.a}$ in bits $3:1$, and the known-bit mask is $\texttt{01110}_2$.

Processing the second constraint gives $\mathit{mask}_{\mathrm{obs}}=\texttt{11100}_2$, with previously seen bits $\mathit{mask}_{\mathrm{seen}}=\texttt{01100}_2$ and new bits $\mathit{mask}_{\mathrm{new}}=\texttt{10000}_2$.
Thus, $\mathrm{D.b}[1:0]$ must agree with the already learned $\mathrm{D.a}[2:1]$, producing the assertion in the figure.
Only $p[4]$ is newly learned, from $\mathrm{D.b}[2]$; the earlier bits are retained.
The final register assignments store $\mathrm{D.b}[2]\mathbin{\texttt{\#\#}}\mathrm{D.a}\mathbin{\texttt{\#\#}}0$ and the known-bit mask $\texttt{11110}_2$, leaving $p[0]$ unknown.

\input{fig/monitor-examples}

\section{Multi-Protocol, Multi-Transaction Compilation}
\label{sec:meta-automata}
The preceding section compiles a protocol into a single-transaction automaton.
We now would like to compose these automata for a finite set of protocols $\mathcal P$  into a single automaton that can drive or monitor an unbounded sequence of transactions with arbitrary interleavings of protocols.
The key step is to
analyze the possible ways multiple transactions can run in parallel and allocate states to each.
We explore all possible configurations of concurrent transactions, effectively ``simulating'' their execution until we recognize recurring patterns. By reusing configurations we have seen before, we can capture unbounded behavior without simulating forever.
For a given protocol, this concurrency behavior is the same for the driver and monitor automaton.
We use a \emph{\meta{}}, which abstracts away driver- or monitor-specific details to perform this analysis, and then lower it back into an ``unbounded driver'' or ``unbounded monitor'' that can handle an unbounded sequence of transactions.

\subsection{Transactions and Configurations}
\label{sec:multiple-transactions}

A \emph{transaction} is one execution of a protocol with its own arguments and registers, e.g., \code|add(5, 2, 7)|.
We write $P_i^n$ for a transaction of protocol $P\in\mathcal P$ currently executing state $i$ of its single-transaction automaton.
The superscript $n$ is its \emph{instance number}, which distinguishes transactions of the same protocol that are currently executing.
We qualify each parameter and register reference by its protocol and instance number.
DUT ports are shared between transactions, so they are not renamed.

A \code|fork| marks the cycle in which a parent transaction launches its child transaction.
In that cycle, the parent transaction executes alongside the child transaction's entry state.
The parent transaction continues independently until it retires, meaning it reaches its exit state and is removed from the list of currently executing transactions.
A transaction is in its \emph{pre-fork phase} before it forks and in its \emph{post-fork phase} from its fork cycle until it retires.

\lang{} requires that each transaction forks exactly once (either explicitly, or implicitly in the exit state), and forbids forking in its first cycle. 
We describe a simple compiler pass to instrument enforcement of these checks in Appendix~\ref{sec:fork-instrumentation}. 
There is at most one transaction in its pre-fork phase, called the \emph{frontier}.
We mark the frontier with an overline.
At a \code|fork|, the child transaction becomes the frontier (since it cannot \code|fork| in its first cycle) as the parent transaction enters its post-fork phase (since it can never \code|fork| again).
At a \code|fork|, any protocol $P \in \mathcal P$ may begin.
In the driver, this decision is made using an input signal from the environment, and in the monitor, this is determined by what protocol's constraints pass on the input signals for that cycle. 

We represent the transactions currently executing as follows.

\begin{definition}[Configuration]
\label{def:configuration}
A \emph{configuration} $C$ is given by the grammar
\[
\begin{aligned}
  F &::= P_i^n \mid F\lor F
  & C &::= P_i^n \mid \overline F \mid C\land P_i^n.
\end{aligned}
\]
A configuration contains at most one frontier and any number of post-fork transactions, with at least one transaction overall.
We mark the frontier with an overline and join it to the other transactions using $\land$, denoting concurrent execution.
Each other transaction is in its post-fork phase and is written $P_i^n$.
The frontier is either a single transaction $\overline{P_i^n}$ or a disjunction of new transactions resulting from a parent's \code|fork|, one for every protocol in $\mathcal P$.
We call the frontier a \emph{choice} in the latter case and use $\lor$ to separate its alternatives.
Under the first-cycle distinguishability assumption (\S\ref{sec:restrictions}), the choice is fully resolved in the first cycle; since a transaction cannot fork in that cycle, the frontier is a single transaction thereafter until it forks.
Thus a frontier choice always includes all protocols, while a frontier that has already made its choice contains just one transaction.
We treat $\lor$ and $\land$ as associative and commutative: grouping and ordering alternatives within a frontier choice or transactions within a configuration does not affect configuration equality.

\end{definition}

\begin{definition}[Configuration automaton]
\label{def:meta-automaton}
A \emph{\meta{}} is a tuple $\mathcal M=(\mathcal C,C_0,\leadsto)$, where $\mathcal C$ is a set of configurations and $C_0\in\mathcal C$ is the initial configuration.
The transition relation $\leadsto$ consists of tuples $(C,g,P,C')$ connecting configurations one clock cycle apart, where $g$ is a Boolean predicate and $P\in\mathcal P\cup\{\emptyset\}$ is the selected protocol, or $\emptyset$ when the transition selects no protocol.
We write $C\overset{g,P}{\leadsto}C'$.
The guard $g$ combines the single-transaction transition guards and fork conditions that permit this successor, with parameter and register references qualified by protocol and instance number.
When $P\ne\emptyset$, the transition selects $P$ from the frontier choice in $C$.
In figures, we draw configurations as ovals in a \meta{}.
\end{definition}

\subsection{Restrictions}
\label{sec:restrictions}
\label{sec:finiteness}
We impose two restrictions so that protocol choices can be resolved and exploration can reuse a finite set of configurations.

\paragraph{First-cycle distinguishability.}
In the driver, the choice of protocol spawned at a \code|fork| is resolved at the end of the first cycle. However, a monitor may observe signals consistent with multiple protocols for several cycles.
To simplify matters, we assume \emph{first-cycle distinguishability}: at most one protocol's entry constraints can pass for the current DUT sample; this will give us one \meta{} that works for both driving and monitoring.
The original \lang{} authors encountered no examples requiring such multi-cycle disambiguation, motivating this simplifying restriction.
A less restrictive extension could retain unresolved frontier choices across cycles, but require the observations to distinguish the candidates before any surviving candidate forks.
This would prevent unresolved choices from multiplying across successive launches and retain the finiteness guarantees established below.
An SMT-based check could encode the candidates' constraints and ask whether distinct protocols can remain consistent with the same signal prefix when a fork is reached, rejecting such cases before compilation.

\paragraph{Bounded post-fork phases.}
For exploration to terminate by reusing recurring configurations as we will see in Section~\ref{sec:steady-states}, we need a bound on the number of simultaneously executing transactions.
However, this example protocol allows unbounded simultaneous transactions:
\begin{lstlisting}[language=paso]
prot p<D: RTL>() {D.in := 0; step(); fork(); D.in := X; while D.out == 0 { step(); } step();}
\end{lstlisting}
If the RTL design keeps \code|D.out| at zero, each transaction launches a child transaction and remains in its post-fork loop.
To fix this problem, we limit how long each transaction can continue after it forks.
This bounds the number of simultaneously executing transactions,
which in turn bounds the number of possible configurations.

A single-transaction automaton has a \emph{bounded post-fork phase} with bound $B$ if every maximal control path after a fork in cycle $t$ reaches the exit state or ends by cycle $t+B$.
We establish this bound by checking that the portion of the control graph reachable after a fork is acyclic.

\begin{proposition}[Finite configurations]
\label{prop:meta-construction-termination}
For a finite family of finite single-transaction automata with bounded post-fork phases, there are finitely many configurations up to instance numbers.
\end{proposition}
\begin{proof}
Let $B$ be the largest post-fork bound in the family.
At most one transaction forks per cycle, since only the frontier can fork and its child cannot fork in its first cycle.
In cycle $t$, any transaction currently executing that has forked must have done so in $[t-B,t]$.
There are therefore at most $B+1$ such transactions and at most one frontier, giving a bound of $B+2$ on simultaneously executing transactions.
Each transaction has finitely many possible protocols and states, and at most $B+2$ instance numbers suffice after renumbering.
Thus there are finitely many configurations up to instance numbers.
\end{proof}

\subsection{Exploring and Reusing Configurations}
\label{sec:steady-states}
Under these restrictions, we construct a \meta{} by exploring which configurations can follow the initial configuration, then repeating this exploration from each configuration discovered.
The initial configuration is a frontier choice among the protocols' entry states.
To find a successor, select one protocol if the frontier is a choice, then advance each transaction currently executing along a transition of its single-transaction automaton.
When the frontier transaction forks, it adds a child as a choice in the frontier.
A transaction reaching its exit state retires and is removed from the configuration.
The transactions that remain form a successor configuration, connected to the preceding configuration by a transition.
Exploring all remaining candidates accounts for the possible future execution of the transactions.
Appendix~\ref{app:meta-construction} gives the formal construction.

When a transaction retires, we remove it from the configuration and can reuse its instance number for a child transaction instead of assigning a fresh instance number (one not previously assigned to a transaction of that protocol during exploration). Consider the family of protocols $\mathcal P$ containing only \code|add| from Figure~\ref{fig:alu-automata}(a).
After the first cycle, each cycle checks one transaction's result while its child transaction starts.
In Figure~\ref{fig:adder-meta}(a), the dotted branches show candidate configurations obtained by assigning the child a fresh instance number.
The solid transitions reuse the retired transaction's instance number, allowing the construction to return to a configuration already explored.
Exploration terminates once every successor of every explored configuration is already represented.
The restrictions in \S\ref{sec:restrictions} guarantee that a finite set of configurations suffices.
\input{fig/adder-meta}

\subsection{Optimization: Reusing Configurations via Renaming}
Reusing retired instance numbers bounds the number of identifiers needed, but can still leave multiple configurations that differ only in their assignment of identifiers to live transactions.
Notice in Figure~\ref{fig:adder-meta}(a) we have two configurations that differ only in instance numbers: each has a parent \code|add| transaction check its result while its child \code|add| drives its inputs.
For a protocol like \code|add| with pipeline depth $d$ (in this case $d=1$), launching a transaction every cycle keeps $n=d+1$ transactions alive, so we need $n$ instance numbers.
Here, the protocol is straight-line, so transactions retire in order; reusing each retiring transaction's instance number for the next launch rotates the numbers through the $n$ pipeline positions, producing $n$ recurring configurations that differ only in numbering. However, more generally, with $n$ live transactions of the same protocol, variable transaction durations can allow different retirement orders, producing up to $n!$ numberings of the same underlying combination of transaction states.

To avoid this extra source of exponential growth in the number of configurations in the \meta{}, we identify configurations as equivalent \emph{up to instance numbers}. To preserve their semantics, we label \meta{} transitions with a \emph{renumbering} that permutes the instance-qualified symbols in $S$. 
At the protocol automaton level, a \textsc{Step} transition (\S\ref{sec:small-step}) applies a permutation to the symbols in the environment $\sigma$, so each surviving transaction's arguments and register values follow its new instance number.

\begin{definition}[Equivalence up to instance numbers]
Two configurations are \emph{equivalent up to instance numbers} if a one-to-one renumbering within each protocol makes their collections of concurrently executing transactions and their frontier choices identical.
A \emph{renumbering} on a transition permutes the instance symbols within each protocol.
An arrow $P^i\to P^j$ carries the arguments and updated registers in slot $i$ of protocol $P$ to slot $j$ in the next configuration.
\end{definition}

In Figure~\ref{fig:adder-meta}(b), $2\to1$ carries the surviving transaction to slot $1$, and $1\to2$ makes the retired transaction's slot available for its child, which is then initialized in slot $2$. 

\input{fig/alu-meta}

In Figure~\ref{fig:alu-meta}, we illustrate the construction for \code|add| and \code|mul|; here transitions occur both on the choice of protocol, as well as the transitions within the single-transaction automata. In Appendix~\ref{app:meta-construction}, we give the formal algorithm that constructs \meta[automata]{}.

\subsection{Generating Unbounded Drivers from Configuration-Automata}
\label{sec:driver-unordered-merge}
The \meta{} gives the possible configurations of concurrently executing transactions.
We now wish to construct a driver from this \meta{}, and later a monitor.
Here, each configuration will become a state of the resulting automaton by combining the actions of the underlying single-transaction automata we built in \S\ref{sec:single-protocol-compilation}.

To construct an unbounded driver, we add an input \code|protocol_choice| to the external-signal set $E$ for selecting the next protocol.
For each transaction $P_i^n$ in a configuration, copy the actions of state $q_i$ from the single-transaction driver for $P$, qualifying parameter and register references by $P$ and $n$.
Discard the \code|fork| actions, which have already been accounted for by the \meta{}.
For a frontier choice, condition the actions for each protocol $P$ on $\texttt{protocol\_choice}=P$.

We then combine the resulting actions by \emph{parallel merge}. This handles two things from \lang{}'s semantics. 
First, a concrete assignment from one transaction takes precedence over another's \code|X|, since releasing a port should not prevent another transaction from driving it.
Second, conflicting concrete assignments are rejected as data races.
Appendix~\ref{app:parallel-merge} gives the formal rules.

To connect up the lowered states, for each $C\overset{g,P}{\leadsto}C'$ with $P\ne\emptyset$, add a step transition between the driver states for $C$ and $C'$ guarded by $g\land(\texttt{protocol\_choice}=P)$.
For $C\overset{g,\emptyset}{\leadsto}C'$, use $g$ alone.

\subsection{Generating Unbounded Monitors from Configuration-Automata}
\label{sec:unbounded-monitors}
We populate the same \meta{} with single-transaction monitor automata using the construction and parallel merge rules from \S\ref{sec:driver-unordered-merge}.
The difference is that protocol selection is not driven by an external signal. Instead, our constraints in the form of \code|assert_eq| actions move onto the transition guards.
For each configuration-automaton edge $e:C\overset{g,P}{\leadsto}C'$, let $\mathcal C_e$ collect the guarded assertions of the currently executing transactions and, when $P\ne\emptyset$, the selected protocol's entry state.
Between the corresponding monitor states, add a step transition guarded by
\[
  g\land
    \bigwedge_{[h]\,\operatorname{assert\_eq}(u,w)\in\mathcal C_e}
      \bigl(\neg h\lor(u=w)\bigr).
\]

\section{Evaluation}

We implement a compiler from \lang{} into protocol automata, reusing the open-source \lang{} frontend.
We also implement a SystemVerilog backend that turns the final automata from Section~\ref{sec:meta-automata} into register-based state machines.
This section demonstrates three use cases for this compiler:
monitoring security properties;
generating ``glue code'' for an HLS compiler;
and efficiently reconstructing transactions from a signal trace.

\subsection{Synthesizable Security Bus Monitor}

Zonta-Roudes et al. \cite{expect_2024, xray_2025} recently identified security vulnerabilities in implementations of the industry-standard AXI on-chip communication protocol.
For example, leaving stale read data on the bus after a transaction completes can expose potentially confidential data to untrusted parties.
We use our compiler to generate efficient in-deployment hardware monitors that check for similar security properties. 
We systematically translate the security properties in that prior work from AXI to Wishbone~\cite{Wishbone:2010}, a similar communication bus supported in \citet{ng_et_al_2026}'s artifact.
We added over 70 assignments and assertions encoding these properties across all 17 protocols in the \lang{} Wishbone specification, covering multiple read and write modes, reset, and idle behavior.
We then compile the protocols to a hardware monitor that checks these security properties.

To evaluate the monitor, we integrate it at the main-RAM Wishbone port of an open-source VexRiscv-based SoC using the LiteX framework~\cite{litex_wishbone}.
The SoC uses the standard VexRiscv RISC-V core, a 32\,KiB boot ROM, 4\,KiB SRAM, and 64\,KiB main RAM, with PCIe and external DDR disabled.
We run Vivado synthesis, placement, and routing targeting an Alveo U250 FPGA at 125\,MHz, keeping the SoC, BIOS, and timing constraints identical across configurations.
The monitor's violation output drives a user LED.

We apply two additional compiler optimizations to reduce monitor resource use.
First, certain Wishbone protocols access successive addresses in \code|for|--\code|in| loops; in the monitor, we can replace the subtractions introduced by constraint normalization (\S\ref{sec:monitor-automata}) with an incrementing register.
Second, because these Wishbone protocols execute transactions serially, registers belonging to different protocols have nonoverlapping live ranges and can share the same storage.

\input{fig/wishbone-evaluation}

Table~\ref{tab:wishbone-evaluation} compares versions of the hardware with and without the monitor.
The optimized monitor adds only 244 LUTs (11.1\%) and 107 flip-flops (6.2\%), with no additional BRAM tiles or DSPs.
It maintains the SoC's 125\,MHz target clock rate, with $+2.847$\,ns setup slack.

\subsection{Calling External Hardware from HLS Designs}
High-level synthesis (HLS) tools compile software-like languages into hardware~\cite{nigam2021calyx,canis2011legup,zhang2008autopilot,pilato2013bambu,putnam2008chimps}.
These compilers typically provide a way to ``call'' into RTL modules.
For example, the Calyx HLS compiler~\cite{nigam2021calyx}
calls primitive SystemVerilog modules using
either Calyx's go-done protocol or a fixed-latency, feed-forward pipeline interface~\cite{kim2024unifying}.
Integrating RTL and HLS therefore requires handwritten adapters.

We use our compiler to automate this process.
We implement a Calyx foreign-function interface (FFI) generator that
translates \lang{} specifications with scalar parameters~\footnote{Calyx's go-done interface requires all input parameters to be supplied in a single cycle. An unbounded sequence cannot be transferred over a fixed set of wires in one cycle.}
into Calyx adapters.
We begin with the single-protocol compilation flow in Figure~\ref{fig:overview}, adapting it to both drive inputs and recover results.
Calyx supplies the input arguments but not the expected result, such as the parameter $s$ in our \code|add| example.
The adapter must \emph{learn} this result from the DUT's outputs and return it to Calyx.
We apply monitor conversion only to \code|assert_eq| actions and leave assignments unchanged.
The resulting automaton drives the DUT's inputs while learning its outputs.

Because each go-done handshake represents a single invocation, the adapter must execute one transaction rather than launch a stream of them.
We therefore require a single-cycle \code|idle| protocol and explicitly restrict each \code|fork| to launch only idle transactions while the invoked transaction finishes.
With this choice fixed, we remove the \code|fork| actions and pad the post-fork portion with idle behavior, preserving legal bus behavior without starting another operation.

We combine the protocols into one adapter by merging their starting states into a shared entry state, which waits for \code|go| to be asserted and starts the protocol selected by \code|op|.
All protocols terminate in a shared \textit{done} state, which pulses \code|done| to signal completion to Calyx.
Figure~\ref{fig:calyx-adapter} illustrates this approach with an open-source FIFO from the Basejump STL library~\cite{taylor2018basejump}, its \lang{} specification, and Calyx code that invokes its push and pop transactions.

\input{fig/calyx-adapter}

\subsection{A Faster Software Reconstructor}

The original \lang{} toolchain includes a \emph{reconstructor},
which infers transaction traces from waveforms.
We reimplement the reconstructor using automata, and the result is vastly more efficient.
The original \lang{} reconstructor uses an exploration approach inspired by dynamic symbolic execution.
Our new reconstructor works by compiling the specification
into a monitor automaton, which we then execute with a fast bytecode interpreter.
The result can make efficient use of memory:
we can statically allocate all storage once at the beginning of execution, avoiding any dynamic heap allocations.
It also avoids rediscovering constraints on every cycle.

\autoref{tab:fast-bi} compares its performance to the original reconstructor
using the Wishbone benchmarks taken from \citet{ng_et_al_2026}:
an OpenCores WB Ethernet core, four configurations of a FuseSoc WB data streaming core,
and an example \texttt{read} transaction waveform from the Wishbone specification.
All wall-clock time measurements were averaged over 10 runs, performed using Hyperfine~\cite{hyperfine} on a server with dual 8-core Intel Xeon E5-2620v4 CPUs.
\vspace{-3pt}
\input{fig/fast-bi-evaluation}

The new reconstructor easily surpasses the original one when
traces are long enough to amortize the compilation overhead.
On long waveforms (e.g., the OpenCores Ethernet waveform containing over 31 million clock cycles), the performance of
our reconstructor achieves over an order of magnitude improvement (see also \autoref{fig:fast-bi-timing-ethmac} in \autoref{sec:fast-bi-ethmac}). However, on short waveforms (e.g., the Wishbone spec example waveform), 
our reconstructor's runtime is dominated by the overhead of compiling protocols to automata, so the original reconstructor wins out.

\section{Related Work}

\paragraph{Type systems for safe hardware composition.}
Recently introduced DSLs such as Filament \cite{filament,lilac} and Anvil~\cite{yu2026anvil,Jhacommunicationsafehardware2027} encode communication specifications in their type systems to ensure safe hardware composition. 
This enables efficient compositional verification of hardware designs that are fully written in their language, but not for interactions with existing RTL components expressed in other languages.
\lang{} specifications, on the other hand, describe how existing RTL components communicate.
With our work, we can use this information to generate correct-by-construction FFI adapters to integrate legacy RTL into HLS compilers.

\paragraph{Runtime monitoring.}
Runtime verification derives monitors from specifications of observed behavior, including temporal logics such as linear temporal logic (LTL) and timed linear temporal logic (TLTL)~\cite{bauer2011runtime}.
LOLA specifies synchronous monitors declaratively through stream equations and identifies specifications whose memory requirements are independent of trace length~\cite{dangelo2005lola}.
Similar to LOLA, we characterize when monitoring an unbounded trace requires only bounded memory (\S\ref{sec:meta-automata}).
RTLola extends stream-based monitoring to real-time properties, and its FPGA compiler uses static resource bounds to produce synthesizable VHDL monitors~\cite{baumeister2019fpga}.

\paragraph{SystemVerilog Assertions (SVA)~\cite{systemverilog} and Property Specification Language (PSL)~\cite{psl}} These LTL and regular--expression-based property languages are integrated with
the hardware description languages SystemVerilog and VHDL. There is an extensive line of work compiling these languages into synthesizable monitors,
either through state machine composition~\cite{pellauer2005synthesis,das2006synthesis}, pattern matching~\cite{wei2024zoomie} or automata-based lowerings~\cite{boule2008automata,yosys_austrochip_2013}.
\lang{} specifications are \emph{imperative} instead of declarative.
Our work is the first to lower an imperative specification language 
to hardware monitors via an automata-based representation.
We also demonstrate synthesizable drivers (used in the Calyx FFI) which cannot directly be generated from SVA or PSL.

\paragraph{High-level synthesis.}
HLS compilers~\cite{cong2011high,nigam2021calyx,canis2011legup,zhang2008autopilot,pilato2013bambu,putnam2008chimps} translate imperative programs into functionally equivalent hardware.
Traditional HLS source languages are timing agnostic; \lang{}, on the other hand, specifies subtle multi-cycle timing behavior and thus requires a different compilation approach.

\section{Conclusion}
A hardware communication specification should be a single source of truth, allowing hardware modules to safely interact with each other.
We show that protocols expressed as simple imperative programs admit formal, automata-based semantics, and we design a compiler that resolves sequential semantics into parallel states and supports unbounded transaction streams in finite space. 
Thus, we enable synthesizable hardware monitors and correct-by-construction FFI adapters for safe hardware composition.

\section{Generative AI Use}
All ideas and text in this paper were human-created.
OpenAI Codex was used to develop experiment scripts, which were then manually evaluated for accuracy. 
When developing the artifact, Claude was used to debug errors that were later fixed by humans.
When preparing this paper, Codex was used to translate drawings into TikZ figures. 
We reviewed all resulting figures and take full responsibility for all content in this paper.

\section{Data Availability Statement}
We have an open-source Rust artifact implementing the compiler detailed in this paper, based on the original \lang{} open-source artifact. 
It is capable of generating all automata shown in this paper, has backends for all targets used in evaluation, and evaluation scripts to reproduce all data shown.
We will submit to artifact evaluation. However, the artifact is not anonymized and evaluation scripts are not fully-automated at this time.

\bibliographystyle{ACM-Reference-Format}
\bibliography{bib/papers}

\newpage

\appendix
\input{appendix}

\end{document}

%% file: fig/driver-monitor.tex
\begin{tikzpicture}[
  x=.72cm,y=.5cm,
  archblock/.style={draw=black!75,fill=black!3,line width=.65pt,
    minimum width=1.6cm,minimum height=.55cm,inner sep=2pt,
    font=\sffamily\small},
  archwire/.style={draw=black!75,line width=.7pt},
  archarrow/.style={archwire,-{Latex[length=1.7mm,width=1.15mm]}},
  archport/.style={font=\sffamily\scriptsize,fill=white,inner sep=1.2pt},
  archtitle/.style={font=\sffamily\small\bfseries,align=center},
  archcell/.style={font=\sffamily\scriptsize,inner sep=0pt},
  archheader/.style={archcell,font=\sffamily\scriptsize\bfseries},
  archjunction/.style={circle,fill=black!75,inner sep=0pt,minimum size=2pt},
  testbench layout/.try
]
\path[use as bounding box] (0,0) rectangle (19.1,5.2);
\draw[black!18,line width=.35pt] (6,.2) -- (6,4.95);
\draw[black!18,line width=.35pt] (12,.2) -- (12,4.95);

\begin{scope}[shift={(6,.1)}]
  \node[archtitle] at (3,4.6) {(b) Monitor};
  \node[archblock] (env1) at (3,3.55) {Env};
  \node[archblock] (mon1) at (3,2.15) {Monitor};
  \node[archblock] (dut1) at (3,.75) {DUT};
  \draw[archwire] (.85,.75) -- (.85,3.55);
  \draw[archwire] (5.15,.75) -- (5.15,3.55);
  \draw[archwire] (env1.west) -- (.85,3.55);
  \node[archjunction] at (.85,3.55) {};
  \node[archjunction] at (5.15,3.55) {};
  \foreach \block/\height in {mon1/2.15,dut1/.75} {
    \draw[archarrow] (.85,\height) -- (\block.west);
    \node[archjunction] at (.85,\height) {};
    \node[archjunction] at (5.15,\height) {};
  }
  \draw[archarrow] (5.15,3.55) -- (env1.east);
  \draw[archarrow] (5.15,2.15) -- (mon1.east);
  \draw[archarrow] (dut1.east) -- (5.15,.75);
  \node[archport] at (1.4,1.18) {in};
  \node[archport] at (4.6,1.18) {out};
\end{scope}

\begin{scope}[shift={(0,.1)}]
  \node[archtitle] at (3,4.6) {(a) Driver};
  \node[archblock] (driver2) at (3,3.1) {Driver};
  \node[archblock] (dut2) at (3,1.4) {DUT};
  \draw[archwire] (.85,1.4) -- (.85,3.1);
  \draw[archwire] (5.15,1.4) -- (5.15,3.1);
  \draw[archarrow] (driver2.west) -- (.85,3.1);
  \draw[archarrow] (.85,1.4) -- (dut2.west);
  \draw[archarrow] (dut2.east) -- (5.15,1.4);
  \draw[archarrow] (5.15,3.1) -- (driver2.east);
  \foreach \height in {1.4,3.1} {
    \node[archjunction] at (.85,\height) {};
    \node[archjunction] at (5.15,\height) {};
  }
  \node[archport] at (1.4,1.83) {in};
  \node[archport] at (4.6,1.83) {out};
\end{scope}

\begin{scope}[shift={(12,.1)}]
  \node[archtitle] at (3.55,4.6) {(c) Interpretation};
  \node[draw=black!75,fill=black!3,rounded corners=1pt,
    minimum width=4.7cm,minimum height=.42cm,inner sep=2pt,
    font=\ttfamily\fontsize{8.5}{10}\selectfont] at (3.55,3.7)
    {\textcolor{cb-blue!75!black}{add(1,2,3)};\quad
     \textcolor{pasoOrange!75!black}{add(7,4,11)}};
  \draw[archarrow] (2.5,3.25) -- (2.5,2.3);
  \node[font=\sffamily\fontsize{8}{9.5}\selectfont,anchor=east,inner sep=1pt]
    at (2.4,2.8) {Driver};
  \draw[archarrow] (4.6,2.3) -- (4.6,3.25);
  \node[font=\sffamily\fontsize{8}{9.5}\selectfont,anchor=west,inner sep=1pt]
    at (4.7,2.8) {Monitor};
  \node[font=\sffamily\fontsize{7.5}{9}\selectfont,anchor=east,inner sep=0pt]
    at (.9,1.95) {Cycle};
  \foreach \x/\cycle in {1.975/1,3.725/2,5.475/3}
    \node[font=\fontsize{8}{9.5}\selectfont,inner sep=0pt] at (\x,1.95) {\cycle};
  \foreach \port/\y/\first/\second/\third/\firstcolor/\secondcolor/\thirdcolor in {
    a/1.4/1/7/0/cb-blue/pasoOrange/black,
    b/.8/2/4/0/cb-blue/pasoOrange/black,
    s/.2/0/3/11/black/cb-blue/pasoOrange} {
    \node[font=\ttfamily\fontsize{8}{9.5}\selectfont,anchor=east,inner sep=0pt]
      at (.9,\y) {\port};
    \foreach \x/\value/\tone in {1.1/\first/\firstcolor,2.85/\second/\secondcolor,4.6/\third/\thirdcolor} {
      \draw[draw=\tone!75!black,fill=\tone!12,line width=.7pt]
        (\x,\y) -- (\x+.1,\y+.25) -- (\x+1.65,\y+.25)
        -- (\x+1.75,\y) -- (\x+1.65,\y-.25) -- (\x+.1,\y-.25) -- cycle;
      \node[font=\ttfamily\fontsize{8}{9.5}\selectfont,inner sep=0pt,text=\tone!75!black]
        at (\x+.875,\y) {\value};
    }
  }
\end{scope}

\end{tikzpicture}

%% file: fig/alu-source.tex
\begin{figure}[t]
\centering
\begin{minipage}{.91\linewidth}
\centering
\begin{minipage}[t]{.305\linewidth}
{\footnotesize\textbf{(a) Depth-1 addition}}
\begin{lstlisting}[language=paso,basicstyle=\ttfamily\fontsize{7}{8.4}\selectfont]
prot add<D: ALU>(
    a: u32, b: u32, s: u32) {
  D.a := a; D.b := b;
  D.op := 0; D.go := 0;
  step();
  D.a := X; D.b := X;
  D.op := X; D.go := X;
  assert_eq(D.s, s);
  fork();
  step();
}
\end{lstlisting}
\end{minipage}\hspace{.03846\linewidth}%
\begin{minipage}[t]{.305\linewidth}
{\footnotesize\textbf{(b) Go--done multiplication}}
\begin{lstlisting}[language=paso,basicstyle=\ttfamily\fontsize{7}{8.4}\selectfont]
prot mul<D: ALU>(
    a: u32, b: u32, s: u32) {
  D.a := a; D.b := b;
  D.op := 1; D.go := 1;
  while !D.done { step(); }
  D.a := X; D.b := X;
  D.go := 0; D.op := X;
  assert_eq(D.s, s);
  step();
}
\end{lstlisting}
\end{minipage}\hspace{.03846\linewidth}%
\begin{minipage}[t]{.305\linewidth}
{\footnotesize\textbf{(c) Example execution}}\par\smallskip
\input{fig/alu-waveform}
\end{minipage}
\end{minipage}
\caption{\lang{} specifications for driving \code|add| and \code|mul| transactions onto the ALU.
In (c), \code|add| starts in cycle 1 and forks in cycle 2, launching \code|mul| while checking the \code|add|'s result; the \code|mul|'s result is available in cycle 3 when \code|done| rises and \code|go| falls combinationally in response.}
\label{fig:alu-source}
\end{figure}
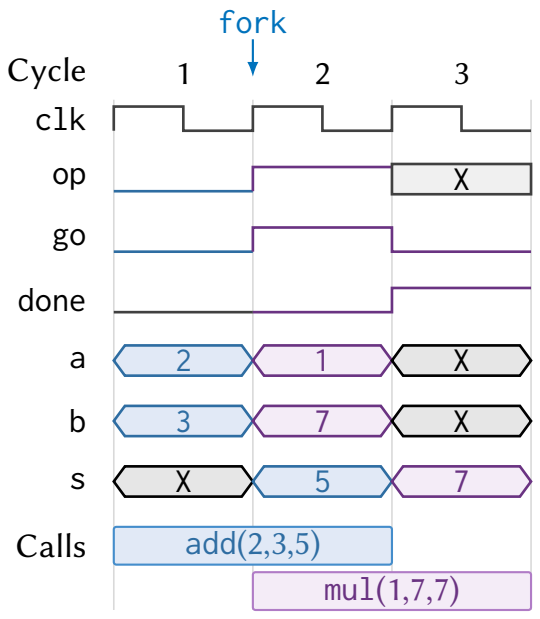

%% file: fig/alu-waveform.tex
\begin{tikzpicture}[
  x=.92cm,y=.5cm,
  wavesignal/.style={draw=black!75,line width=.5pt},
  wavetext/.style={font=\fontsize{6.5}{7.5}\selectfont,inner sep=1pt},
  waveport/.style={wavetext,font=\ttfamily\fontsize{6.5}{7.5}\selectfont,anchor=east}
]
  \foreach \x in {1,2,3,4}
    \draw[black!15,line width=.3pt] (\x,.25) -- (\x,-6.5);
  \node[wavetext,anchor=east] at (.85,.6) {Cycle};
  \foreach \x/\cycle in {1.5/1,2.5/2,3.5/3}
    \node[wavetext] at (\x,.6) {\cycle};
  \node[wavetext,text=cb-blue] at (2,1.28) {\texttt{fork}};
  \draw[-{Latex[length=1mm]},cb-blue,line width=.4pt] (2,1.05) -- (2,.55);
  \foreach \port/\y in {clk/0,op/-.8,go/-1.6,done/-2.4,a/-3.2,b/-4,s/-4.8}
    \node[waveport] at (.85,\y) {\port};
  \draw[wavesignal] (1,-.16) -- (1,.16) -- (1.5,.16) -- (1.5,-.16)
    -- (2,-.16) -- (2,.16) -- (2.5,.16) -- (2.5,-.16)
    -- (3,-.16) -- (3,.16) -- (3.5,.16) -- (3.5,-.16) -- (4,-.16);
  \draw[wavesignal,draw=cb-blue!75!black] (1,-.96) -- (2,-.96);
  \draw[wavesignal,draw=pasoPurple!75!black] (2,-.96) -- (2,-.64) -- (3,-.64);
  \fill[black!6] (3,-1) rectangle (4,-.6);
  \draw[wavesignal] (3,-1) rectangle (4,-.6);
  \node[wavetext] at (3.5,-.8) {\texttt{X}};
  \draw[wavesignal,draw=cb-blue!75!black] (1,-1.76) -- (2,-1.76);
  \draw[wavesignal,draw=pasoPurple!75!black] (2,-1.76) -- (2,-1.44)
    -- (3,-1.44) -- (3,-1.76) -- (4,-1.76);
  \draw[wavesignal] (1,-2.56) -- (2,-2.56);
  \draw[wavesignal,draw=pasoPurple!75!black] (2,-2.56) -- (3,-2.56) -- (3,-2.24) -- (4,-2.24);
  \foreach \y/\first/\second/\third/\firstcolor/\secondcolor/\thirdcolor in {
    -3.2/2/1/X/cb-blue/pasoPurple/black,
    -4/3/7/X/cb-blue/pasoPurple/black,
    -4.8/X/5/7/black/cb-blue/pasoPurple} {
    \foreach \x/\value/\tone in {1/\first/\firstcolor,2/\second/\secondcolor,3/\third/\thirdcolor} {
      \draw[wavesignal,draw=\tone!75!black,fill=\tone!10]
        (\x,\y) -- (\x+.08,\y+.2) -- (\x+.92,\y+.2)
        -- (\x+1,\y) -- (\x+.92,\y-.2) -- (\x+.08,\y-.2) -- cycle;
      \node[wavetext,text=\tone!75!black] at (\x+.5,\y) {\texttt{\value}};
    }
  }
  \node[wavetext,anchor=east] at (.85,-5.65) {Calls};
  \draw[draw=cb-blue!70,fill=cb-blue!10,line width=.4pt,rounded corners=.5pt]
    (1,-5.9) rectangle (3,-5.4);
  \node[wavetext,text=cb-blue!75!black] at (2,-5.65) {\texttt{add}(2,3,5)};
  \draw[draw=pasoPurple!70,fill=pasoPurple!10,line width=.4pt,rounded corners=.5pt]
    (2,-6.5) rectangle (4,-6);
  \node[wavetext,text=pasoPurple!75!black] at (3,-6.25) {\texttt{mul}(1,7,7)};
\end{tikzpicture}

%% file: fig/alu-unaligned.tex
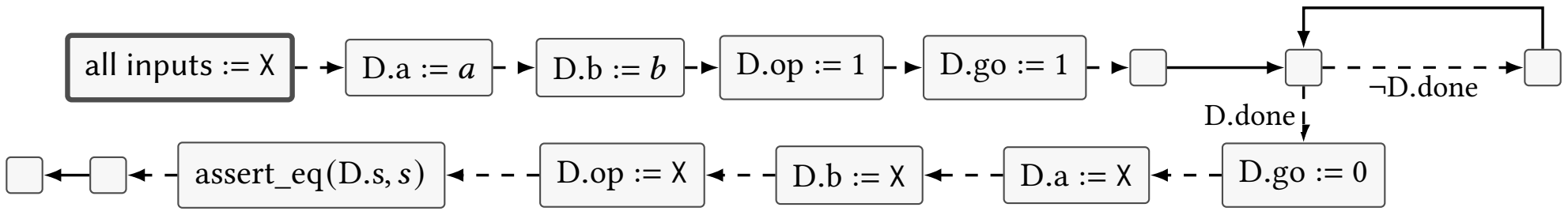
\begin{figure}[t]
\centering
\begingroup
\input{fig/alu-styles}
\begin{tikzpicture}
  \node[alustate,line width=1.2pt] (init) at (0,0) {all inputs $:=\texttt{X}$};
  \node[alustate] (a) at (2,0) {$\alup{a}:=a$};
  \node[alustate] (b) at (3.6,0) {$\alup{b}:=b$};
  \node[alustate] (op) at (5.2,0) {$\alup{op}:=1$};
  \node[alustate] (go) at (6.9,0) {$\alup{go}:=1$};
  \node[aluempty] (step) at (8.1,0) {};
  \foreach \from/\to in {init/a,a/b,b/op,op/go,go/step}
    \draw[alunonstep] (\from) -- (\to);
  \node[aluempty] (head) at (9.4,0) {};
  \draw[aluedge] (step) -- (head);
  \node[aluempty] (wait) at (11.4,0) {};
  \draw[alunonstep] (head) -- node[alulabel,below] {$\neg\alup{done}$} (wait);
  \draw[aluedge] (wait.north) -- (11.4,.5) -- (9.4,.5) -- (head.north);
  \node[alustate] (releasego) at (9.4,-.9) {$\alup{go}:=0$};
  \node[alustate] (relea) at (7.5,-.9) {$\alup{a}:=\texttt{X}$};
  \node[alustate] (releb) at (5.6,-.9) {$\alup{b}:=\texttt{X}$};
  \node[alustate] (releop) at (3.7,-.9) {$\alup{op}:=\texttt{X}$};
  \node[alustate] (check) at (1.1,-.9) {$\aluck{\alup{s}}{s}$};
  \draw[alunonstep] (head) -- node[alulabel,left] {$\alup{done}$} (releasego);
  \foreach \from/\to in {releasego/relea,relea/releb,releb/releop,releop/check}
    \draw[alunonstep] (\from) -- (\to);
  \node[aluempty] (exit) at (-.6,-.9) {};
  \node[aluempty] (done) at (-1.3,-.9) {};
  \draw[alunonstep] (check) -- (exit);
  \draw[aluedge] (exit) -- (done);
\end{tikzpicture}
\endgroup
\caption{Unaligned protocol automaton for \code|mul| from Fig.~\ref{fig:alu-source}. A bold border indicates the initial state. Solid arrows are step edges (which advance time by one cycle); dashed arrows are nonstep edges (which do not).}
\label{fig:alu-unaligned}
\end{figure}

%% file: fig/alu-styles.tex
\newcommand{\alup}[1]{\mathrm{D.#1}}
\newcommand{\aluck}[2]{\operatorname{assert\_eq}(#1,#2)}
\tikzset{
  alustate/.style={draw=black!70,fill=black!3,rounded corners=1pt,
    align=left,inner sep=4pt,font=\fontsize{8}{9.5}\selectfont},
  aluempty/.style={alustate,minimum width=3mm,minimum height=3mm,inner sep=0pt},
  aluedge/.style={-{Latex[length=1.5mm]},line width=.65pt},
  alunonstep/.style={aluedge,dashed},
  alulabel/.style={font=\scriptsize,fill=white,inner sep=1.5pt,align=center},
  aluid/.style={font=\scriptsize,above=2pt}
}

%% file: fig/alu-individual-automata.tex
\begin{figure}[t]
\centering
\begingroup
\input{fig/alu-styles}
\tikzset{alustate/.append style={font=\fontsize{7.5}{9}\selectfont,inner sep=3pt}}
\input{fig/alu-drivers}
\input{fig/alu-monitors}
\begin{minipage}[t]{.32\linewidth}
\centering\vspace{0pt}
{\bfseries (a) \code|add| driver}\par\smallskip
\aluAddDriver
\par\medskip
{\bfseries (c) \code|add| monitor}\par\smallskip
\aluAddMonitor
\end{minipage}\hfill
\begin{minipage}[t]{.335\linewidth}
\centering\vspace{0pt}
{\bfseries (b) \code|mul| driver}\par\smallskip
\aluMulDriver
\end{minipage}\hfill
\begin{minipage}[t]{.315\linewidth}
\centering\vspace{0pt}
{\bfseries (d) \code|mul| monitor}\par\smallskip
\aluMulMonitor
\end{minipage}
\endgroup
\caption{Clock-aligned drivers and monitors. Each node has an integer ID, which will become relevant in \S\ref{sec:meta-automata}.}
\label{fig:alu-automata}
\label{fig:alu-drivers}
\label{fig:alu-monitors}
\end{figure}
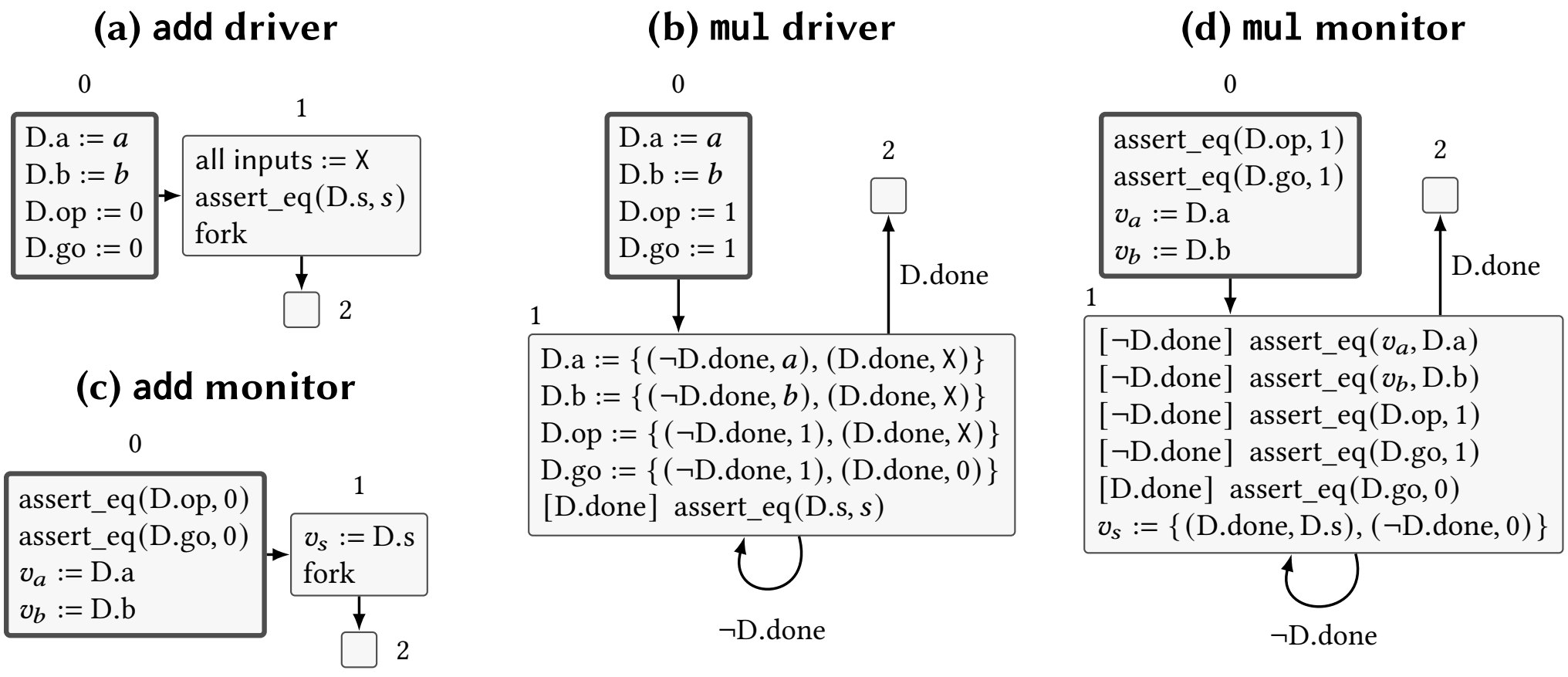

%% file: fig/alu-drivers.tex
\newcommand{\aluAddDriver}{%
\begin{tikzpicture}
  \node[alustate,line width=1.2pt] (a0) at (0,0) {$\alup{a}:=a$\\$\alup{b}:=b$\\$\alup{op}:=0$\\$\alup{go}:=0$};
  \node[alustate,right=2mm of a0] (a1) {all inputs $:=\texttt{X}$\\$\aluck{\alup{s}}{s}$\\$\mathrm{fork}$};
  \node[alustate,font=\scriptsize,inner sep=2pt,minimum width=3mm,minimum height=3mm,below=3mm of a1] (a2) {};
  \node[font=\scriptsize,above=1pt of a0] {$0$};
  \node[font=\scriptsize,above=1pt of a1] {$1$};
  \node[font=\scriptsize,right=1pt of a2] {$2$};
  \draw[aluedge] (a0.east) -- (a1.west);
  \draw[aluedge] (a1.south) -- (a2.north);
\end{tikzpicture}
}
\newcommand{\aluMulDriver}{%
\begin{tikzpicture}
  \node[alustate,line width=1.2pt] (m0) at (-.8,2.05) {$\alup{a}:=a$\\$\alup{b}:=b$\\$\alup{op}:=1$\\$\alup{go}:=1$};
  \node[alustate] (m1) at (0,0) {
    $\alup{a}:=\{(\neg\alup{done},a),(\alup{done},\texttt{X})\}$\\
    $\alup{b}:=\{(\neg\alup{done},b),(\alup{done},\texttt{X})\}$\\
    $\alup{op}:=\{(\neg\alup{done},1),(\alup{done},\texttt{X})\}$\\
    $\alup{go}:=\{(\neg\alup{done},1),(\alup{done},0)\}$\\
    $[\alup{done}]\;\aluck{\alup{s}}{s}$};
  \node[alustate,font=\scriptsize,inner sep=2pt,minimum width=3mm,minimum height=3mm] (m2) at (1,2.05) {};
  \node[font=\scriptsize,above=1pt of m0] {$0$};
  \node[font=\scriptsize,anchor=south west,inner sep=0pt] at ([yshift=2pt]m1.north west) {$1$};
  \node[font=\scriptsize,above=1pt of m2] {$2$};
  \draw[aluedge] (m0.south) -- ([xshift=-.8cm]m1.north);
  \draw[aluedge] ([xshift=1cm]m1.north) --
    node[alulabel,right=1pt] {$\alup{done}$} (m2.south);
  \draw[aluedge] (m1) edge[loop below,min distance=6mm]
    node[alulabel,fill=none,below=2mm] {$\neg\alup{done}$} (m1);
\end{tikzpicture}
}

%% file: fig/alu-monitors.tex
\newcommand{\aluAddMonitor}{%
\begin{tikzpicture}
  \node[alustate,line width=1.2pt] (a0) at (0,0) {$\aluck{\alup{op}}{0}$\\$\aluck{\alup{go}}{0}$\\$v_a:=\alup{a}$\\$v_b:=\alup{b}$};
  \node[alustate,right=2mm of a0] (a1) {$v_s:=\alup{s}$\\$\mathrm{fork}$};
  \node[alustate,font=\scriptsize,inner sep=2pt,minimum width=3mm,minimum height=3mm,below=3mm of a1] (a2) {};
  \node[font=\scriptsize,above=1pt of a0] {$0$};
  \node[font=\scriptsize,above=1pt of a1] {$1$};
  \node[font=\scriptsize,right=1pt of a2] {$2$};
  \draw[aluedge] (a0.east) -- (a1.west);
  \draw[aluedge] (a1.south) -- (a2.north);
\end{tikzpicture}
}
\newcommand{\aluMulMonitor}{%
\begin{tikzpicture}
  \node[alustate,line width=1.2pt] (m0) at (-.8,2.05) {$\aluck{\alup{op}}{1}$\\$\aluck{\alup{go}}{1}$\\$v_a:=\alup{a}$\\$v_b:=\alup{b}$};
  \node[alustate] (m1) at (0,0) {
    $[\neg\alup{done}]\;\aluck{v_a}{\alup{a}}$\\
    $[\neg\alup{done}]\;\aluck{v_b}{\alup{b}}$\\
    $[\neg\alup{done}]\;\aluck{\alup{op}}{1}$\\
    $[\neg\alup{done}]\;\aluck{\alup{go}}{1}$\\
    $[\alup{done}]\;\aluck{\alup{go}}{0}$\\
    $v_s:=\{(\alup{done},\alup{s}),(\neg\alup{done},0)\}$};
  \node[alustate,font=\scriptsize,inner sep=2pt,minimum width=3mm,minimum height=3mm] (m2) at (1,2.05) {};
  \node[font=\scriptsize,above=1pt of m0] {$0$};
  \node[font=\scriptsize,anchor=south west,inner sep=0pt] at ([yshift=2pt]m1.north west) {$1$};
  \node[font=\scriptsize,above=1pt of m2] {$2$};
  \draw[aluedge] (m0.south) -- ([xshift=-.8cm]m1.north);
  \draw[aluedge] ([xshift=1cm]m1.north) --
    node[alulabel,right=1pt] {$\alup{done}$} (m2.south);
  \draw[aluedge] (m1) edge[loop below,min distance=6mm]
    node[alulabel,fill=none,below=1mm] {$\neg\alup{done}$} (m1);
\end{tikzpicture}
}

%% file: fig/alu-unbounded-monitor.tex
\begin{figure}[t]
\centering
\begingroup
\input{fig/alu-styles}
\begin{tikzpicture}[
  aluscope/.style={draw=black!70,fill=white,line width=.5pt,
    font=\scriptsize,inner xsep=3pt,inner ysep=2pt,anchor=north east},
  aluactions/.style={font=\fontsize{8}{9.5}\selectfont,align=left,anchor=west},
  aludivider/.style={draw=black!55,dashed,line width=.5pt},
  alulabel/.append style={fill=none}
]
  \begin{scope}[yshift=-.375cm]
  \node[alustate,draw=none,fill=none,minimum width=2.1cm,minimum height=2.4cm,inner sep=0pt]
    (entry) at (0,.6) {};
  \begin{scope}
    \clip[rounded corners=1pt] (-1.05,-.6) rectangle (1.05,1.8);
    \fill[cb-blue!10] (-1.05,.6) rectangle (1.05,1.8);
    \fill[pasoPurple!10] (-1.05,-.6) rectangle (1.05,.6);
  \end{scope}
  \draw[black!70,line width=1.2pt,rounded corners=1pt] (-1.05,-.6) rectangle (1.05,1.8);
  \draw[aludivider] (-1.05,.6) -- (1.05,.6);
  \node[aluscope,draw=cb-blue,fill=cb-blue!15,text=cb-blue!70!black] at (entry.north east) {$\texttt{Add}^1$};
  \node[aluscope,draw=pasoPurple,fill=pasoPurple!15,text=pasoPurple!80!black]
    at ([yshift=-1.2cm]entry.north east) {$\texttt{Mul}^1$};
  \node[aluactions] at (-.9,1.2) {
    $v_a:=\alup{a}$\\$v_b:=\alup{b}$\\$v_s:=0$};
  \node[aluactions] at (-.9,0) {
    $v_a:=\alup{a}$\\$v_b:=\alup{b}$\\$v_s:=0$};

  \end{scope}

  \node[alustate,draw=none,fill=none,minimum width=2.1cm,minimum height=3.15cm,inner sep=0pt]
    (add) at (3.6,.225) {};
  \begin{scope}
    \clip[rounded corners=1pt] (2.55,-1.35) rectangle (4.65,1.8);
    \fill[cb-blue!10] (2.55,1.05) rectangle (4.65,1.8);
    \fill[pasoGreen!12] (2.55,-.15) rectangle (4.65,1.05);
    \fill[pasoPurple!10] (2.55,-1.35) rectangle (4.65,-.15);
  \end{scope}
  \draw[black!70,line width=.65pt,rounded corners=1pt] (2.55,-1.35) rectangle (4.65,1.8);
  \draw[aludivider] (2.55,1.05) -- (4.65,1.05);
  \draw[aludivider] (2.55,-.15) -- (4.65,-.15);
  \node[aluscope,draw=cb-blue,fill=cb-blue!15,text=cb-blue!70!black] at (add.north east) {$\texttt{Add}^1$};
  \node[aluscope,draw=pasoGreen!75!black,fill=pasoGreen!20,text=pasoGreen!85!black]
    at ([yshift=-.75cm]add.north east) {$\texttt{Add}^2$};
  \node[aluscope,draw=pasoPurple,fill=pasoPurple!15,text=pasoPurple!80!black]
    at ([yshift=-1.95cm]add.north east) {$\texttt{Mul}^1$};
  \node[aluactions] at (2.7,1.425) {$v_s:=\alup{s}$};
  \node[aluactions] at (2.7,.45) {
    $v_a:=\alup{a}$\\$v_b:=\alup{b}$\\$v_s:=0$};
  \node[aluactions] at (2.7,-.75) {
    $v_a:=\alup{a}$\\$v_b:=\alup{b}$\\$v_s:=0$};

  \node[alustate,fill=pasoPurple!10,minimum width=2.8cm,minimum height=1.15cm,inner sep=0pt]
    (mul) at (8.4,.225) {};
  \node[aluscope,draw=pasoPurple,fill=pasoPurple!15,text=pasoPurple!80!black] at (mul.north east) {$\texttt{Mul}^1$};
  \node[aluactions] at (7.12,.1) {
    $v_s:=\{(\alup{done},\alup{s}),$\\
    $\phantom{v_s:=\{}(\neg\alup{done},0)\}$};
  \node[font=\scriptsize\bfseries,align=center,above=2pt] at (entry.north) {
    $\texttt{Add}^1$ or $\texttt{Mul}^1$ Starts};
  \node[font=\scriptsize\bfseries,align=center,above=2pt] at (add.north) {
    $\texttt{Add}^1$ Completes,\\
    $\texttt{Add}^2$ or $\texttt{Mul}^1$ Starts};
  \node[font=\scriptsize\bfseries,align=center,above=2pt] at (mul.north) {
    $\texttt{Mul}^1$ Waiting\\for Done};
  \draw[aluedge] (entry.east) -- node[alulabel,above=2mm] {
    \textit{Choose \texttt{Add}}\\
    $(\alup{op}=0)$\\$\land\;(\alup{go}=0)$} (add.west);
  \draw[aluedge] (entry.south east)
    .. controls (1.6,-1.8) and (2.7,-1.85) .. (3.6,-1.85)
    .. controls (5.5,-1.85) and (6.5,-1.6) .. (mul.south west);
  \node[alulabel] at (1.8,-.55) {
    \textit{Choose \texttt{Mul}}\\
    $(\alup{op}=1)$\\$\land\;(\alup{go}=1)$};
  \draw[aluedge] ([yshift=1.1cm]add.east)
    .. controls +(0.75,.2) and +(0.75,-.2) ..
    ([yshift=.4cm]add.east);
  \node[alulabel] at (6.15,1.55) {
    \textit{Choose \texttt{Add}}\\
    \textit{$\texttt{Add}^2$ becomes $\texttt{Add}^1$}\\
    $(\alup{op}=0)$\\$\land\;(\alup{go}=0)$};
  \draw[aluedge] (add.east) --
    node[alulabel,anchor=north west,xshift=1mm,yshift=-2mm,pos=0] {
      \textit{Choose \texttt{Mul}}\\
      $(\alup{op}=1)$\\$\land\;(\alup{go}=1)$} (mul.west);
  \draw[aluedge] ([xshift=.47cm]mul.south)
    .. controls (8.87,-1.95) and (6.2,-2.25) .. (3.6,-2.25)
    .. controls (1,-2.25) and (-1.2,-2.05) .. (entry.south west);
  \node[alulabel,anchor=west] at (8.7,-1.15) {
    \textit{\texttt{Mul} Completes}\\
    $\alup{done}\land(\alup{go}=0)$};
  \draw[aluedge] ([yshift=.3cm]mul.east)
    .. controls +(0.75,.2) and +(0.75,-.2) ..
    ([yshift=-.3cm]mul.east);
  \node[alulabel] at (11.45,.225) {
    \textit{Continue Waiting}\\
    $\neg\alup{done}$\\
    $\land\;(\alup{a}=v_a)$\\
    $\land\;(\alup{b}=v_b)$\\
    $\land\;(\alup{op}=1)$\\
    $\land\;(\alup{go}=1)$};
\end{tikzpicture}
\endgroup
\caption{An unbounded monitor for both protocols.
Dashed dividers separate register scopes, identified by the corner tags; $\texttt{Add}^1$ and $\texttt{Add}^2$ are successive additions, and $\texttt{Mul}^1$ is a multiplication. If no edge is taken, there is an implicit edge to a self-looping failing state.}
\label{fig:alu-unbounded-monitor}
\end{figure}
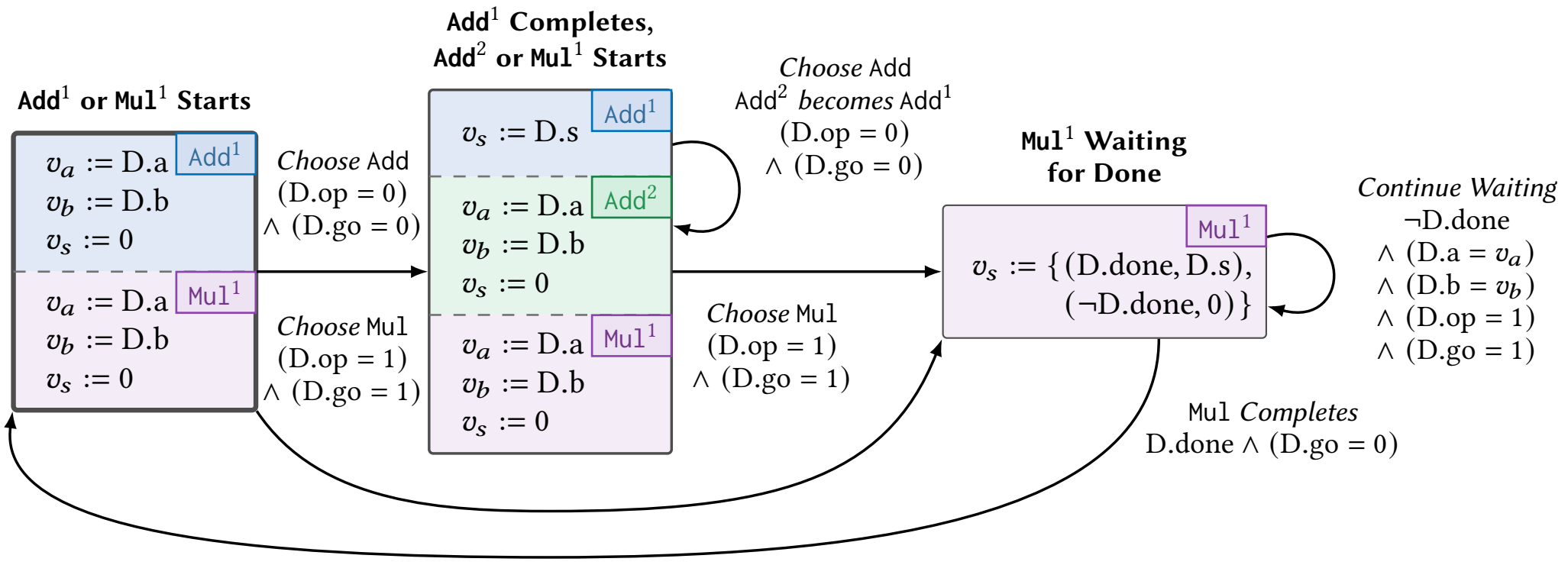

%% file: fig/ast-to-ir.tex
\begin{figure}[t]
\centering
\begingroup
\fontsize{9}{10.5}\selectfont
\input{fig/ast-to-ir-styles}
\begin{tikzpicture}[
  irvertex/.append style={minimum width=1cm,minimum height=6.5mm,inner sep=2pt},
  iraction/.append style={minimum width=1cm,minimum height=6.5mm,inner sep=2pt},
  irbubble/.append style={minimum width=1cm,minimum height=6.5mm,inner sep=2pt},
  irsource/.style={anchor=base west,inner sep=0pt,font=\fontsize{8.5}{10}\selectfont},
  irheading/.style={anchor=west,inner sep=0pt,font=\bfseries}
]
  \node[irheading] at (0,0) {Atomic statements};
  \node[irsource] at (0,-.55) {\code|p := e|};
  \node[iraction,anchor=west] (assign) at (0,-1.15) {$p\!:=\!e$};
  \draw[irnonstep] (assign.east) -- (2.2,-1.15);

  \node[irsource] at (2.8,-.55) {\code|assert_eq(e1, e2)|};
  \node[iraction,anchor=west] (assert) at (2.8,-1.15) {$\operatorname{assert\_eq}(e_1,e_2)$};
  \draw[irnonstep] (assert.east) -- (6.4,-1.15);

  \node[irsource] at (7,-.55) {\code|fork()|};
  \node[iraction,anchor=west] (fork) at (7,-1.15) {$\mathrm{fork}$};
  \draw[irnonstep] (fork.east) -- (9,-1.15);

  \node[irsource] at (9.6,-.55) {\code|step()|};
  \node[irvertex,anchor=west] (step) at (9.6,-1.15) {};
  \node[irvertex] (restore) at (11.7,-1.15) {};
  \draw[irstep] (step) -- (restore);
  \draw[irnonstep] (restore.east) -- (12.9,-1.15);

  \node[irheading] at (0,-1.95) {Recursive statements};
  \begin{scope}[yshift=-1.5mm]
  \node[irsource] at (0,-2.35) {$S_1$\code|; |$S_2$};
  \node[irbubble,anchor=west] (s1) at (0,-3.35) {$S_1$};
  \node[irbubble] (s2) at (1.9,-3.35) {$S_2$};
  \draw[irnonstep] (s1) -- (s2);
  \draw[irnonstep] (s2.east) -- (2.85,-3.35);

  \node[irsource] at (3.1,-2.35) {\code|if g { |$\mathcal T$\code| } else { |$\mathcal F$\code| }|};
  \node[irvertex,anchor=west] (if) at (3.1,-3.35) {};
  \node[irtrue] (t) at (5.1,-2.95) {$\mathcal T$};
  \node[irfalse] (f) at (5.1,-3.9) {$\mathcal F$};
  \node[irjoin] (ifjoin) at (6.8,-3.35) {};
  \draw[irtrueedge] (if) -- node[irlabel,above=1.5pt,text=blue!65!black] {$g$} (t);
  \draw[irfalseedge] (if) -- node[irlabel,below=3.5pt,text=orange!75!black] {$\neg g$} (f);
  \draw[irtrueedge] (t) -- (ifjoin);
  \draw[irfalseedge] (f) -- (ifjoin);
  \draw[irnonstep] (ifjoin.east) -- (7.7,-3.35);

  \node[irsource] at (8,-2.35) {\code|while g { |$\mathcal B$\code| }|};
  \node[irvertex,anchor=west] (while) at (8,-3.35) {};
  \node[irtrue] (loopbody) at (10.3,-3.9) {$\mathcal B$};
  \node[irexits] (whilejoin) at (12,-3.35) {};
  \draw[irfalseedge] (while) -- node[irlabel,above,text=orange!75!black] {$\neg g$} (whilejoin);
  \draw[irnonstep] (whilejoin.east) -- (12.9,-3.35);
  \draw[irtrueedge] (while) -- node[irlabel,below=2pt,text=blue!65!black] {$g$} (loopbody);
  \draw[irtrueedge] (loopbody.south) -- ++(0,-.2) -| (while.south);
  \end{scope}

  \node[irheading] at (0,-4.95) {Top-level lowering};
  \node[irsource] at (0,-5.65) {\code|prot name<D: RTL>(params) { |$\mathcal B$\code| }|};
  \node[iraction,very thick,minimum width=2.55cm,anchor=west] (entry) at (6.8,-5.6) {all inputs $:=\texttt X$};
  \node[irbubble] (body) at (10.45,-5.6) {$\mathcal B$};
  \node[irvertex] (exit) at (12.4,-5.6) {};
  \draw[irnonstep] (entry) -- (body);
  \draw[irstep] (body) -- (exit);
\end{tikzpicture}
\endgroup
\caption{Syntax-directed lowering of statements and complete protocols to the initial automata.
Calligraphic letters denote statement blocks $S_1;\ldots;S_n$, and dotted boxes denote a recursively lowered subgraph.}
\label{fig:ast-to-ir}
\end{figure}
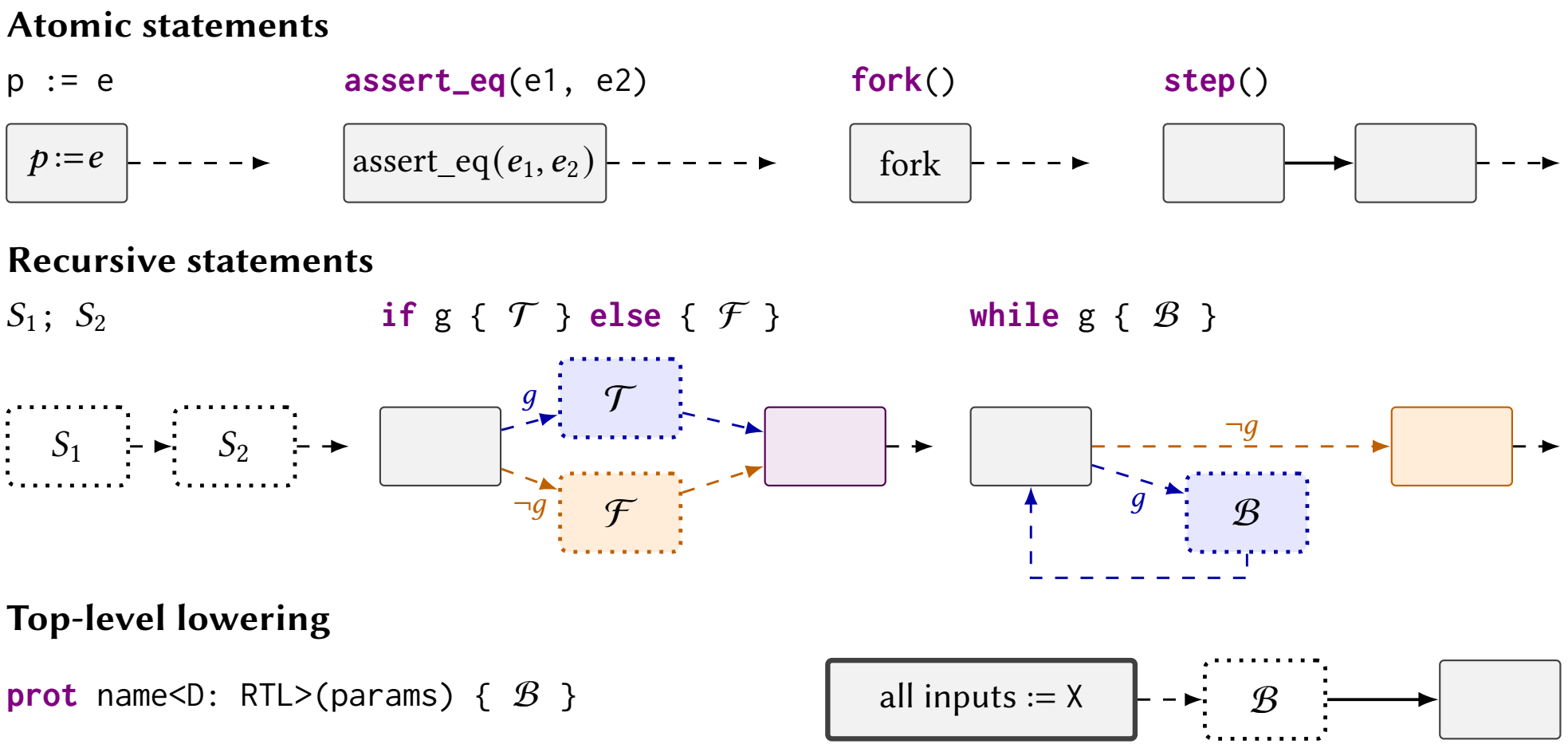

%% file: fig/ast-to-ir-styles.tex
\tikzset{
  irvertex/.style={draw=black!75, fill=black!5, rounded corners=1pt,
                   rectangle, minimum width=7mm, minimum height=5mm,
                   inner sep=1.5pt},
  iraction/.style={irvertex, minimum width=2.55cm,
                   minimum height=8mm, align=center,
                   font=\fontsize{8.5}{10}\selectfont},
  irbubble/.style={draw, rounded corners=2pt, dotted, line width=.9pt, align=center,
                   inner sep=2.5pt, minimum width=7mm, minimum height=7mm},
  irtrue/.style={irbubble, draw=blue!65!black, fill=blue!10},
  irfalse/.style={irbubble, draw=orange!75!black, fill=orange!15},
  irjoin/.style={irvertex, draw=violet!60!black, fill=violet!10},
  irexits/.style={irvertex, draw=orange!75!black, fill=orange!15},
  iredge/.style={-{Latex[length=1.5mm]}, line width=.55pt},
  irnonstep/.style={iredge, dashed},
  irstep/.style={-{Latex[length=1.7mm]}, line width=.7pt},
  irtrueedge/.style={irnonstep, draw=blue!65!black},
  irfalseedge/.style={irnonstep, draw=orange!75!black},
  irlabel/.style={font=\fontsize{7}{8.5}\selectfont,
                  inner sep=1pt},
  irnodelabel/.style={font=\fontsize{8}{9}\selectfont,
                      text=black!70, inner sep=0pt},
}

%% file: fig/adder-holdover.tex
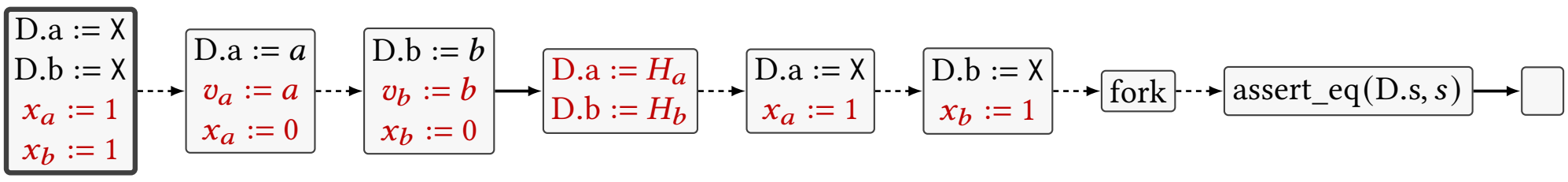
\begin{figure}[t]
\centering
\begingroup
\tikzset{
  ahnode/.style={draw=black!75,fill=black!3,rounded corners=1pt,
    align=center,inner sep=2pt,font=\fontsize{8}{9.5}\selectfont},
  ahempty/.style={ahnode,minimum width=3.5mm,minimum height=4mm,inner sep=0pt},
  ahedge/.style={-{Latex[length=1.2mm]},line width=.5pt},
  ahnonstep/.style={ahedge,dash pattern=on 1.3pt off 1.1pt},
  ahstep/.style={ahedge,line width=.7pt},
}
\begin{tikzpicture}
  \node[ahnode,line width=1.2pt,anchor=west] (entry) at (0,0) {$\mathrm{D.a}:=\texttt X$\\$\mathrm{D.b}:=\texttt X$\\
    $\color{red!75!black}x_a:=1$\\$\color{red!75!black}x_b:=1$};
  \node[ahnode,right=4mm of entry] (a) {$\mathrm{D.a}:=a$\\
    $\color{red!75!black}v_a:=a$\\$\color{red!75!black}x_a:=0$};
  \node[ahnode,right=4mm of a] (b) {$\mathrm{D.b}:=b$\\
    $\color{red!75!black}v_b:=b$\\$\color{red!75!black}x_b:=0$};
  \node[ahnode,right=4mm of b] (restore1)
    {$\color{red!75!black}\mathrm{D.a}:=H_a$\\$\color{red!75!black}\mathrm{D.b}:=H_b$};
  \node[ahnode,right=4mm of restore1] (releasea) {$\mathrm{D.a}:=\texttt X$\\
    $\color{red!75!black}x_a:=1$};
  \node[ahnode,right=4mm of releasea] (releaseb) {$\mathrm{D.b}:=\texttt X$\\
    $\color{red!75!black}x_b:=1$};
\node[ahnode,right=4mm of releaseb] (fork) {$\mathrm{fork}$};
\node[ahnode,right=4mm of fork] (check) {$\operatorname{assert\_eq}(\mathrm{D.s},s)$};
\node[ahempty,right=4mm of check] (exit) {};
\draw[ahnonstep] (entry) -- (a);
\draw[ahnonstep] (a) -- (b);
\draw[ahstep] (b) -- (restore1);
\draw[ahnonstep] (restore1) -- (releasea);
\draw[ahnonstep] (releasea) -- (releaseb);
\draw[ahnonstep] (releaseb) -- (fork);
\draw[ahnonstep] (fork) -- (check);
\draw[ahstep] (check) -- (exit);
\end{tikzpicture}
\endgroup
\caption{The \code|add| protocol instrumented to persist assignments, with $\mathcal{I}(v_a)=\mathcal{I}(v_b)=0$ and $\mathcal{I}(x_a)=\mathcal{I}(x_b)=1$.
Here $H_p\triangleq\{(x_p,\texttt{X}),(\neg x_p,v_p)\}$ for $p\in\{a,b\}$ for brevity. \textcolor{red!75!black}{Red} marks instrumented actions.}
\label{fig:adder-holdover}
\end{figure}

%% file: fig/wire-substitution.tex
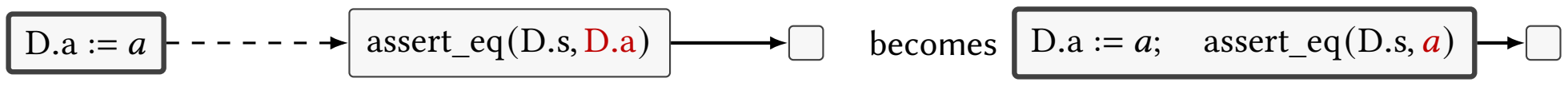
\begin{figure}[t]
\centering
\begingroup
\small
\begin{minipage}{\linewidth}
\raggedright
\textbf{DUT (Verilog):}\quad\lstinline[language=Verilog,basicstyle=\ttfamily\fontsize{8}{9.5}\selectfont]|module Wire(input [31:0] a, output [31:0] s); assign s = a; endmodule|\par
\vspace{3pt}
\textbf{Protocol:}\quad\lstinline[language=paso]|prot wire<D: Wire>(a: u32) { D.a := a; assert_eq(D.s, D.a); step(); }|\par
\vspace{6pt}
\end{minipage}
\begin{tikzpicture}[
  wsnode/.style={draw=black!75,fill=black!3,rounded corners=1pt,
    align=center,inner sep=4pt,font=\fontsize{8.5}{10}\selectfont},
  wsedge/.style={-{Latex[length=1.5mm]},line width=.6pt},
  wsnonstep/.style={wsedge,dashed},
  wsstep/.style={wsedge,line width=.8pt},
]
\node[wsnode,line width=1.2pt] (drive) at (0,0) {$\mathrm{D.a}:=a$};
\node[wsnode] (check) at (3.5,0) {$\operatorname{assert\_eq}(\mathrm{D.s},{\color{red!75!black}\mathrm{D.a}})$};
\node[wsnode] (beforedone) at (5.95,0) {};
\draw[wsnonstep] (drive) -- (check);
\draw[wsstep] (check) -- (beforedone);
\node[font=\fontsize{8.5}{10}\selectfont,inner sep=0pt] at (7,0) {becomes};
\node[wsnode,line width=1.2pt,anchor=west] (merged) at (7.65,0) {$\mathrm{D.a}:=a;\quad\operatorname{assert\_eq}(\mathrm{D.s},{\color{red!75!black}a})$};
\node[wsnode,right=4mm of merged] (afterdone) {};
\draw[wsstep] (merged) -- (afterdone);
\end{tikzpicture}
\endgroup
\caption{Contracting a combinational-wire protocol; \textcolor{red!75!black}{red} highlights the substituted read.}
\label{fig:wire-substitution}
\end{figure}

%% file: fig/sequential-merge.tex
\begin{table}[t]
\caption{Sequential action merge rules for a later state reached under path condition $g$.}
\label{tab:sequential-merge}
\begingroup
\papertablestyle

\newcommand{\gact}[2]{[#1]\,#2}
\newcommand{\failact}{\mathrm{fail}}

\parbox{\columnwidth}{First substitute $p\mapsto V_p$ in all later expressions using the summaries accumulated from the cycle entry.
Then apply the first matching rule below.}
\par\smallskip

\begin{tabular}{@{}p{.28\columnwidth}p{.25\columnwidth}p{.39\columnwidth}@{}}
\toprule
Earlier action & Later action (rewritten) & Actions after the merge \\
\midrule
$\gact{g_1}{\failact}$
  & $\gact{g_2}{\failact}$
  & $\gact{g_1\lor(g\land g_2)}{\failact}$ \\
$\gact{g_1}{\mathrm{fork}}$
  & $\gact{g_2}{\mathrm{fork}}$
  & $\gact{g_1\lor(g\land g_2)}{\mathrm{fork}}$,
    merge $\gact{g_1\land g\land g_2}{\failact}$ \\
$p:=V_p$
  & $p:=V'_p$
  & $p:=V_p\mathbin{\blacktriangleright}(g\cdot V'_p)$ \\
No matching earlier action
  & $b$
  & Insert $g\cdot b$. \\
$a$
  & No matching later action
  & Retain $a$. \\
\bottomrule
\end{tabular}

\endgroup

\end{table}

%% file: fig/isolate-params.tex
\begin{table}[t]
\caption{Constraint normalization rules. Each reduction inherits the guard $g$.}
\label{tab:constraint-normalization}
\papertablestyle
\begin{tabular}{@{}p{.29\columnwidth}p{.20\columnwidth}p{.43\columnwidth}@{}}
\toprule
Constraint & Condition & Reduction \\
\midrule
\multicolumn{3}{@{}l}{\textit{Recursive reductions}} \\
\addlinespace[2pt]
$V_p[h:l]=\mathit{word}$
  & $g'$ parameter-free.
  & $[g\land g']\;e[h:l]=\mathit{word}$ for each
    $(g',e)\in V_p$. \\
$(e[u:v])[h:l]=\mathit{word}$
  & Well-typed ranges.
  & $e[v+h:v+l]=\mathit{word}$ \\
$(e_1\mathbin{\texttt{\#\#}}e_2)[h:l]=\mathit{word}$
  & $h<r$
  & $e_2[h:l]=\mathit{word}$ \\
$(e_1\mathbin{\texttt{\#\#}}e_2)[h:l]=\mathit{word}$
  & $l\ge r$
  & $e_1[h-r:l-r]=\mathit{word}$ \\
$(e_1\mathbin{\texttt{\#\#}}e_2)[h:l]=\mathit{word}$
  & $l<r\le h$
  & $\begin{aligned}[t]
      e_1[h-r:0]&=\mathit{word}[h-l:r-l],\\
      e_2[r-1:l]&=\mathit{word}[r-l-1:0].
    \end{aligned}$ \\
$(e+c)[h:l]=\mathit{word}$
  & $c$ parameter-free; $l=0$ or $c[l-1:0]=0$.
  & $e[h:l]=\mathit{word}-c[h:l]$ \\
\midrule
\multicolumn{3}{@{}l}{\textit{Terminal cases}} \\
\addlinespace[2pt]
$a[h:l]=\mathit{word}$
  & $a$ is a parameter.
  & Record this normalized constraint. \\
$\texttt{X}=\mathit{word}$
  &
  & No observation or check. \\
$c[h:l]=\mathit{word}$
  & $c$ parameter-free.
  & Emit $[g]\;\operatorname{assert\_eq}(c[h:l],\mathit{word})$. \\
\bottomrule
\end{tabular}
\end{table}

%% file: fig/monitor-examples.tex
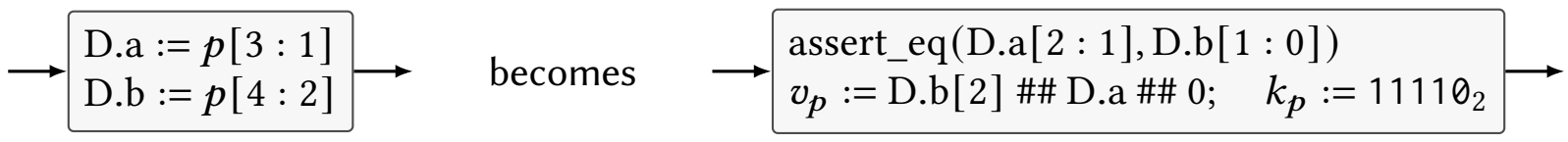
\begin{figure}[H]
\centering
\begingroup
\begin{tikzpicture}[
  monexample/.style={draw=black!70, fill=black!3, rounded corners=1pt,
    align=left, inner sep=3pt, font=\fontsize{8}{9.5}\selectfont},
  monstep/.style={-{Latex[length=1.2mm]},line width=.6pt},
]
  \node[monexample] (driver) at (0,0) {
    $\mathrm{D.a}:=p[3:1]$\\
    $\mathrm{D.b}:=p[4:2]$};
  \node[font=\fontsize{8}{9.5}\selectfont,inner sep=0pt,right=9mm of driver] (becomes) {becomes};
  \node[monexample,right=9mm of becomes] (monitor) {
    $\operatorname{assert\_eq}(\mathrm{D.a}[2:1],\mathrm{D.b}[1:0])$\\
    $v_p:=\mathrm{D.b}[2]\mathbin{\texttt{\#\#}}\mathrm{D.a}\mathbin{\texttt{\#\#}}0;\quad
    k_p:=\texttt{11110}_2$};
  \foreach \state in {driver,monitor} {
    \draw[monstep] ([xshift=-4mm]\state.west) -- (\state.west);
    \draw[monstep] (\state.east) -- ([xshift=4mm]\state.east);
  }
\end{tikzpicture}
\endgroup
\caption{Overlapping observations and the resulting monitor state.}
\label{fig:monitor-examples}
\end{figure}

%% file: fig/adder-meta.tex
\begin{figure}[t]
\centering
\begin{minipage}[c]{.55\linewidth}
\centering
\begin{tikzpicture}[
  metanode/.style={draw=black!70,fill=black!3,ellipse,
    font=\footnotesize,align=center,inner sep=2pt,minimum height=6mm},
  metaedge/.style={-{Latex[length=1.5mm]},line width=.65pt},
]
  \node[metanode,line width=1.2pt,minimum width=1cm] (p) at (0,0) {$\overline{\texttt{add}_0^1}$};
  \node[metanode,minimum width=2.2cm] (s0) at (2,0) {$\texttt{add}_1^1\land \overline{\texttt{add}_0^2}$};
  \node[metanode,minimum width=2.2cm] (s1) at (4.7,0) {$\texttt{add}_1^2\land \overline{\texttt{add}_0^1}$};
  \node[metanode,densely dotted,draw=black!45,fill=none,text=black!65]
    (candidate) at (2,-1.15)
    {$\texttt{add}_1^2\land\overline{\texttt{add}_0^3}$};
  \draw[metaedge,densely dotted,draw=black!45]
    (s0.south) -- (candidate);
  \node[metanode,densely dotted,draw=black!45,fill=none,text=black!65]
    (candidate2) at (4.7,-1.15)
    {$\texttt{add}_1^1\land\overline{\texttt{add}_0^3}$};
  \draw[metaedge,densely dotted,draw=black!45]
    (s1.south) -- (candidate2.north);
  \draw[metaedge] (p) -- (s0);
  \draw[metaedge] (s0) -- (s1);
  \draw[metaedge] (s1.north) to[out=90,in=90,looseness=.4] (s0.north);
\end{tikzpicture}
\par\smallskip{\bfseries (a)}
\end{minipage}\hfill
\begin{minipage}[c]{.43\linewidth}
\centering
\begin{tikzpicture}[
  metanode/.style={draw=black!70,fill=black!3,ellipse,
    font=\footnotesize,align=center,inner sep=2pt,minimum height=6mm},
  metaedge/.style={-{Latex[length=1.5mm]},line width=.65pt},
  metalabel/.style={font=\scriptsize,inner sep=2pt,align=left},
]
  \node[metanode,line width=1.2pt,minimum width=1.2cm] (p) at (0,0) {$\overline{\texttt{add}_0^1}$};
  \node[metanode,minimum width=2.3cm] (s) at (2.15,0) {$\texttt{add}_1^1\land \overline{\texttt{add}_0^2}$};
  \draw[metaedge] (p) -- (s);
  \draw[metaedge] (s.20) .. controls +(1.1,.45) and +(1.1,-.45) ..
    node[metalabel,right,pos=.5] {$2\to1$\\$1\to2$} (s.-20);
\end{tikzpicture}
\par\smallskip{\bfseries (b)}
\end{minipage}
\caption{The \meta[automata]{} for \code|add| using (a) configuration equality and (b) configuration equivalence up to instance numbers.
The gray dotted configurations in (a) show successors we could generate with fresh instance numbers, but the solid transitions are preferred because they reuse configurations.
In (b), we compress the two states that are equivalent up to renumbering into a single state.}
\label{fig:adder-meta}
\end{figure}
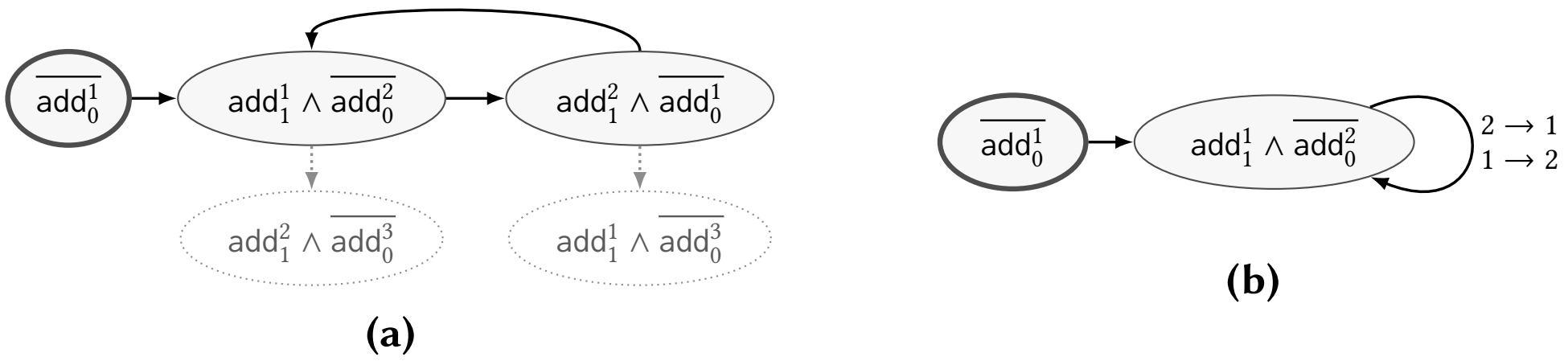

%% file: fig/alu-meta.tex
\begin{figure}[!t]
\centering
\begingroup
\tikzset{
  buildconfig/.style={ellipse,draw=black!65,fill=black!3,
    font=\fontsize{8}{9.5}\selectfont,align=center,inner sep=3pt,
    minimum height=7mm},
  buildedge/.style={-{Latex[length=1.4mm]},line width=.65pt,draw=black!55,rounded corners=2pt},
  buildlabel/.style={font=\fontsize{7}{8}\selectfont,inner sep=2pt},
  buildtitle/.style={font=\fontsize{7}{8}\selectfont\bfseries,align=center,inner sep=2pt},
  buildnew/.style={draw=red,line width=1pt},
}
\newcommand{\buildpanel}[1]{%
\begin{tikzpicture}
  \tikzset{added1/.style={},added2/.style={},added3/.style={},
    added4/.style={},added5/.style={},added6/.style={},
    added#1/.style={buildnew}}
  \ifnum#1=0
    \path[use as bounding box] (-2.05,-2.45) rectangle (.5,.75);
    \node[anchor=north west,font=\bfseries\small] at (-2.05,.55) {(1)};
  \else
    \ifnum#1=1
      \path[use as bounding box] (-2.55,-2.45) rectangle (.6,.75);
    \else
      \ifnum#1=2
        \path[use as bounding box] (-2.55,-2.45) rectangle (3,.75);
      \else
        \path[use as bounding box] (-2.55,-2.3) rectangle (3.5,.9);
      \fi
    \fi
    \ifnum#1<3
      \def\paneltop{.55}
    \else
      \def\paneltop{.6}
    \fi
    \node[anchor=north west,font=\bfseries\small] at (-2.55,\paneltop)
      {(\number\numexpr#1+1\relax)};
  \fi
  \ifnum#1<3
    \def\entryheight{-.25}
  \else
    \def\entryheight{-.1}
  \fi
  \ifnum#1<3
    \def\bodyheight{-1.45}
  \else
    \def\bodyheight{-1.3}
  \fi
  \node[buildconfig,line width=1.2pt] (initial) at (-.8,\entryheight)
    {$\overline{\textcolor{cb-blue!70!black}{\texttt{add}_0^1}\lor\textcolor{pasoPurple!80!black}{\texttt{mul}_0^1}}$};
  \ifnum#1=0
    \node[buildtitle,below=2pt] at (initial.south)
      {$\texttt{Add}^1$ or $\texttt{Mul}^1$ Starts};
  \else
  \node[buildconfig] (add) at (-.8,\bodyheight)
    {$\textcolor{cb-blue!70!black}{\texttt{add}_1^1}\land\overline{\textcolor{pasoGreen!85!black}{\texttt{add}_0^2}\lor\textcolor{pasoPurple!80!black}{\texttt{mul}_0^1}}$};
  \ifnum#1>1
    \ifnum#1=2
      \def\mulposition{2.2}
    \else
      \def\mulposition{2.7}
    \fi
    \node[buildconfig] (mul) at (\mulposition,\bodyheight)
      {$\overline{\textcolor{pasoPurple!80!black}{\texttt{mul}_1^1}}$};
  \fi
  \ifnum#1=1
    \node[buildtitle,below=2pt] at (add.south)
      {$\texttt{Add}^1$ Completes,\\
       $\texttt{Add}^2$ or $\texttt{Mul}^1$ Starts};
  \fi
  \ifnum#1=2
    \node[buildtitle,below=2pt] at (mul.south)
      {$\texttt{Mul}^1$\\Waiting for Done};
  \fi
  \draw[buildedge,added1] (initial.south) --
    node[buildlabel,left] {\texttt{add}} (add.north);
  \ifnum#1>1
    \draw[buildedge,added2] (initial.east) --
      node[buildlabel,above] {\texttt{mul}} (mul.north |- initial.east)
      -- (mul.north);
  \fi
  \ifnum#1>2
    \draw[buildedge,added3] (add.south east)
      .. controls +(.3,-.4) and +(-.3,-.4) ..
      node[buildlabel,below,align=center,font=\fontsize{7}{8}\selectfont] {
        \texttt{add}}
      (add.south west);
  \fi
  \ifnum#1>3
    \draw[buildedge,added4] (add) --
      node[buildlabel,above] {\texttt{mul}} (mul);
  \fi
  \ifnum#1>4
    \draw[buildedge,added5] (mul.south east)
      .. controls +(.2,-.4) and +(-.2,-.4) ..
      node[buildlabel,below,align=center,font=\fontsize{7}{8}\selectfont] {
        $\neg\texttt{D.done}$}
      (mul.south west);
  \fi
  \ifnum#1>5
    \draw[buildedge,buildnew] (mul.east) -- (3.5,\bodyheight)
      -- (3.5,.65) --
      node[buildlabel,above] {$\texttt{D.done}$}
      (initial.north |- 0,.65) -- (initial.north);
  \fi
  \fi
\end{tikzpicture}%
}
\begin{tabular}{@{}c@{\hspace{2mm}}c@{\hspace{6mm}}c@{}}
\buildpanel{0} & \buildpanel{1} & \buildpanel{2}
\end{tabular}
\par\smallskip
\begin{tabular}{@{}c@{\hspace{2mm}}c@{}}
\buildpanel{3} & \buildpanel{4}\\
\buildpanel{5} & \buildpanel{6}
\end{tabular}
\endgroup
\caption{Construction of the \meta{} for \code|add| and \code|mul|.
(1) The initial frontier choice has no transitions yet.
(2) Selecting \code|add| creates a configuration where the \code|add| checks its result while a new \code|add| or \code|mul| begins, and (3) selecting \code|mul| transitions into the \code|mul| waiting for \code|D.done| to go high.
In (4), selecting another \code|add| retires the parent and reuses the configuration by swapping slots 1 and 2 (elided in the diagram). 
In (5), selecting \code|mul| retires the \code|add| and reuses the waiting configuration.
In (6), a waiting \code|mul| remains waiting, and (7) \code|mul| retires, returning to the initial configuration.
}
\label{fig:alu-meta}
\end{figure}
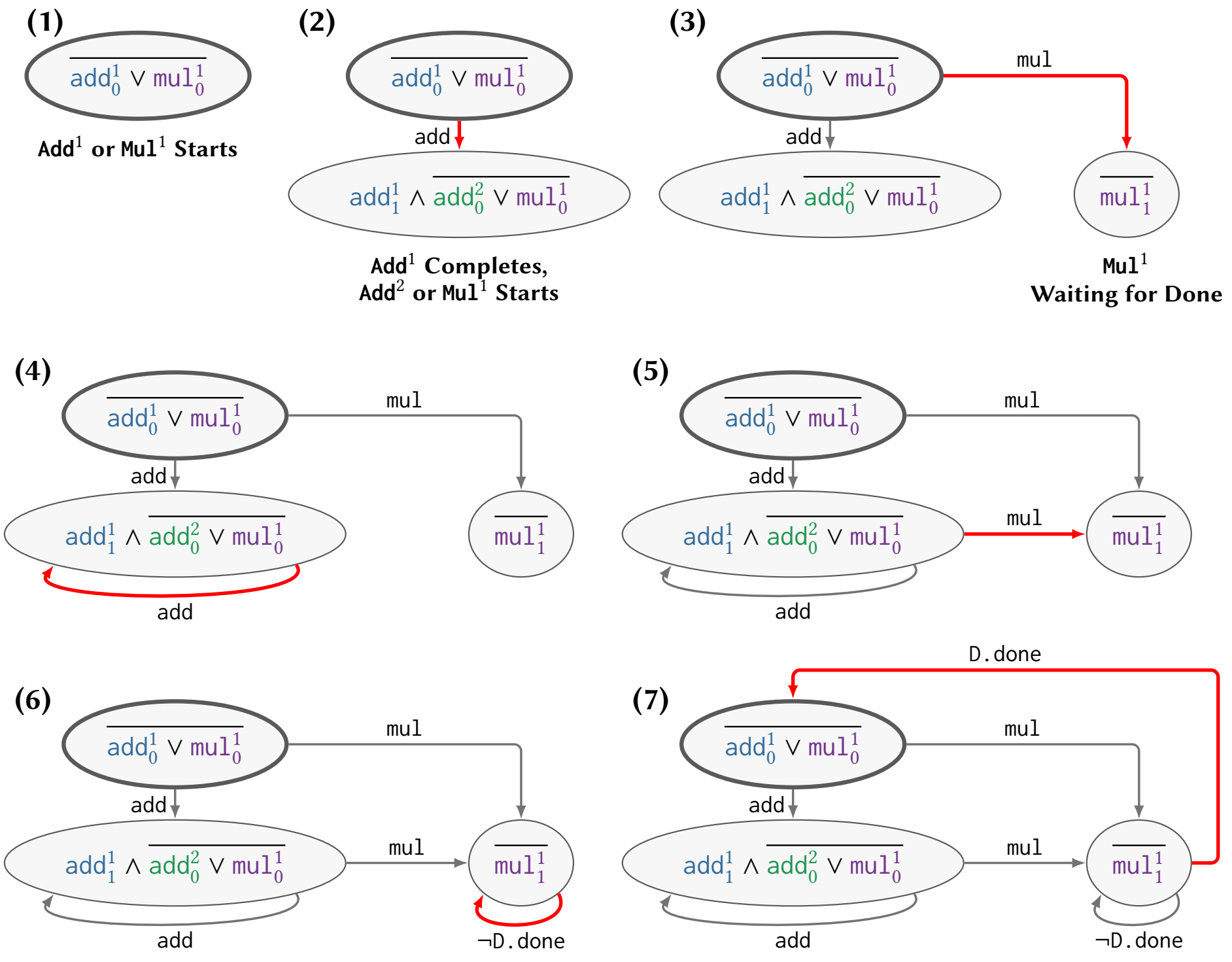

%% file: fig/wishbone-evaluation.tex
\begin{table}[t]
\caption{Whole-SoC resource use and timing with the Wishbone monitor.
Percentages are the overhead relative to the unmonitored baseline, and WNS stands for worst negative slack.}
\label{tab:wishbone-evaluation}
\papertablestyle
\begin{tabular}{@{}lrrr@{}}
\toprule
 & Baseline & Generated & Optimized \\
\midrule
LUTs & 2,205 & 2,591 (+17.5\%) & 2,449 (+11.1\%) \\
Flip-flops & 1,735 & 1,973 (+13.7\%) & 1,842 (+6.2\%) \\
BRAM tiles & 38 & 38 & 38 \\
DSPs & 4 & 4 & 4 \\
125\,MHz setup slack & $+3.272$\,ns & $+2.760$\,ns & $+2.847$\,ns \\
Overall WNS & $+1.186$\,ns & $+1.235$\,ns & $+1.298$\,ns \\
\bottomrule
\end{tabular}
\end{table}

%% file: fig/calyx-adapter.tex
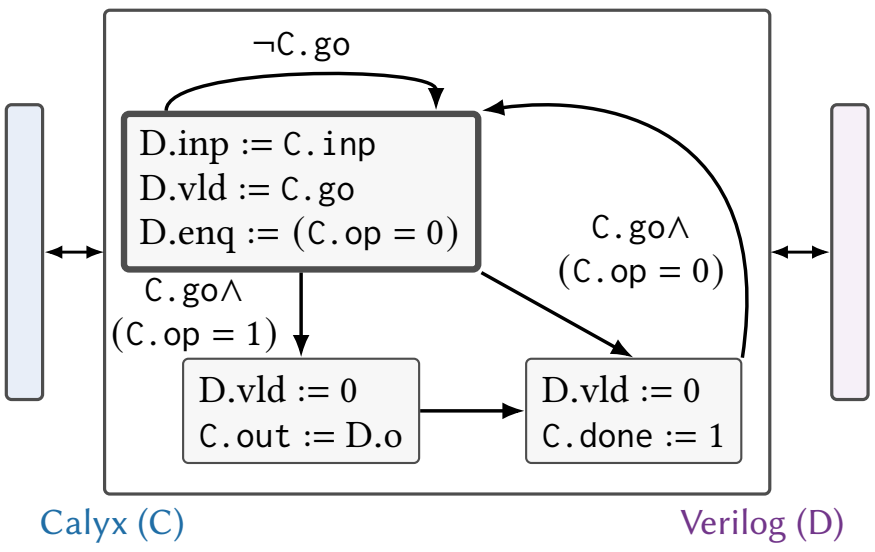
\begin{figure}[t]
\centering

\begin{minipage}[c]{.29\linewidth}
\centering
\begin{lstlisting}[language=paso,basicstyle=\ttfamily\fontsize{7}{8.4}\selectfont,aboveskip=0pt,belowskip=0pt]
prot push<D: Fifo>(inp: u32) {
  D.enq := 1'b1;
  D.vld := 1'b1; D.inp := inp;
  step();
}
prot pop<D: Fifo>(out: u32) {
  D.enq := 1'b0;
  D.vld := 1'b1; 
  step(); fork();
  assert_eq(D.o,out); step();
}
\end{lstlisting}
\end{minipage}\hfill
\newsavebox{\calyxadaptercodebox}
\begin{minipage}[c]{.27\linewidth}
\centering
\begingroup
\setlength{\fboxsep}{2pt}
\begin{lrbox}{\calyxadaptercodebox}
\begin{minipage}{\dimexpr\linewidth-2\fboxsep\relax}
\begin{lstlisting}[language=calyx,basicstyle=\ttfamily\fontsize{7}{8.4}\selectfont,backgroundcolor={},aboveskip=0pt,belowskip=0pt]
repeat 10 { // push 10 vals
  invoke fifo(op=1'b0,
    inp=count_32.out)();
  incr;
}
invoke count(in=4'd0)();
repeat 10 { // pop 10 vals
  invoke fifo(op=1'b1)();
  invoke use(in=fifo.out)();
  incr;
}
\end{lstlisting}
\end{minipage}
\end{lrbox}
\colorbox{cb-blue!8}{\usebox{\calyxadaptercodebox}}
\endgroup
\end{minipage}\hfill
\raisebox{-1.25mm}{%
\begin{minipage}[c]{.415\linewidth}
\centering
\input{fig/calyx-adapter-automaton}
\end{minipage}}
\par\vspace{2pt}
{\footnotesize\bfseries
\makebox[.29\linewidth]{(a) \lang{} Spec}\hfill
\makebox[.27\linewidth]{(b) Calyx HLS code}\hfill
\makebox[.415\linewidth]{(c) Generated adapter}}

\caption{A FIFO adapter generated from (a) its \lang{} specification and invoked by (b) Calyx code. (c) The adapter drives the Verilog FIFO, learns \code|C.out| from \code|D.o|, and pulses \code|C.done| on completion.
}
\label{fig:calyx-adapter}
\end{figure}

%% file: fig/calyx-adapter-automaton.tex
\begingroup
\input{fig/alu-styles}
\begin{tikzpicture}[
  adapterbox/.style={draw=black!70,rounded corners=1pt,line width=.6pt},
  adapterwire/.style={{Latex[length=1.2mm]}-{Latex[length=1.2mm]},line width=.55pt},
  adapterstate/.style={alustate,font=\fontsize{7}{8.4}\selectfont,inner sep=3pt},
  adapterguard/.style={alulabel,font=\fontsize{7}{8.4}\selectfont,fill=none,align=center},
  adaptername/.style={font=\fontsize{7}{8.4}\selectfont,anchor=north,inner sep=1pt},
]
  \draw[adapterbox] (-2.2,-1.45) rectangle (2.2,1.75);
  \node[adapterbox,fill=cb-blue!8,minimum width=.24cm,minimum height=1.95cm]
    (calyx) at (-2.73,.15) {};
  \node[adapterbox,fill=pasoPurple!8,minimum width=.24cm,minimum height=1.95cm]
    (fifo) at (2.73,.15) {};
  \node[adaptername,text=cb-blue!80!black] at (-2.15,-1.5) {Calyx (C)};
  \node[adaptername,text=pasoPurple!80!black] at (2.15,-1.5) {Verilog (D)};
  \draw[adapterwire] (calyx.east) -- (-2.2,.15);
  \draw[adapterwire] (2.2,.15) -- (fifo.west);

  \begin{scope}[yshift=-2mm]
  \node[adapterstate,line width=1.2pt] (entry) at (-.9,.75) {
    $\alup{inp}:=\texttt{C.inp}$\\
    $\alup{vld}:=\texttt{C.go}$\\
    $\alup{enq}:=(\texttt{C.op}=0)$};
  \node[adapterstate] (pop) at (-.9,-.7) {
    $\alup{vld}:=0$\\
    $\texttt{C.out}:=\alup{o}$};
  \node[adapterstate] (done) at (1.3,-.7) {
    $\alup{vld}:=0$\\
    $\texttt{C.done}:=1$};

  \draw[aluedge] (entry.south) --
    node[adapterguard,left=2pt] {$\texttt{C.go}\land$\\$(\texttt{C.op}=1)$} (pop.north);
  \draw[aluedge] (pop.east) -- (done.west);
  \draw[aluedge] (entry.south east) -- (done.north);
  \node[adapterguard] at (1.35,.35) {$\texttt{C.go}\land$\\$(\texttt{C.op}=0)$};
  \draw[aluedge] (done.north east)
    .. controls (2.25,1.2) and (1.1,1.4) .. (entry.north east);
  \draw[aluedge] ([xshift=3mm]entry.north west)
    .. controls +(0,3mm) and +(0,3mm) ..
    node[adapterguard,above=1pt,pos=.5] {$\neg\texttt{C.go}$}
    ([xshift=-3mm]entry.north east);
  \end{scope}
\end{tikzpicture}
\endgroup

%% file: fig/fast-bi-evaluation.tex
\begin{table}[h]
\caption{
  Wall-clock time of Ng et al.'s and our reconstructors on various Wishbone (WB) benchmarks taken from Ng et al.'s evaluation.
  Following Ng et al., ``All Protocols'' indicates the full Wishbone specification containing 17 protocols
  and ``Reduced Protocols'' uses the subset of protocols that occur in the waveform.
}
\label{tab:fast-bi}
\papertablestyle
\begin{tabular}{@{}lrrrrr@{}}
\toprule
 & & \multicolumn{2}{c}{Ng et al.'s Reconstructor} & \multicolumn{2}{c}{Our Reconstructor} \\
\cmidrule(lr){3-4} \cmidrule(l){5-6}
Benchmark & Total Cycles & All Protocols & Reduced Protocols & All Protocols & Reduced Protocols \\
\midrule
OpenCores Ethernet & 31,663,326 & 615.68\,s (1.00$\times$) & 236.39\,s (2.60$\times$) & 41.50\,s (14.84$\times$) & 21.24\,s (28.99$\times$) \\
WB Stream (Read Data) & 201,903 & 8.643\,s (1.00$\times$) & 1.114\,s (7.76$\times$) & 0.935\,s (9.24$\times$) & 0.208\,s (41.47$\times$) \\
WB Stream (Read Control) & 201,903 & 10.305\,s (1.00$\times$) & 1.233\,s (8.36$\times$) & 0.925\,s (11.14$\times$) & 0.149\,s (69.07$\times$) \\
WB Stream (Write Data) & 91,165 & 1.500\,s (1.00$\times$) & 0.330\,s (4.54$\times$) & 0.765\,s (1.96$\times$) & 0.148\,s (10.13$\times$) \\
WB Stream (Write Control) & 91,165 & 4.523\,s (1.00$\times$) & 0.561\,s (8.07$\times$) & 0.798\,s (5.67$\times$) & 0.091\,s (49.63$\times$) \\
WB Spec Example Waveform & 11 & 0.008\,s (1.00$\times$) & 0.003\,s (2.68$\times$) & 0.544\,s (0.01$\times$) & 0.015\,s (0.51$\times$) \\
\bottomrule
\end{tabular}
\end{table}

%% file: appendix.tex
\section{Enforcing Single Forks}
\label{sec:fork-instrumentation}
\lang{} has two special restrictions on \code|fork|.
First, a protocol can explicitly \code|fork| at most once \cite{ng_et_al_2026}.
To check this property at runtime, we introduce a one-bit register $h$, initially false ($\mathcal{I}(h) = 0$), whose value denotes whether an earlier state in a cycle-aligned automaton has already \code|fork|ed.
At a state with a \code|fork| guarded by $g$, we update the register $h$ by taking its disjunction with $g$, assigning $h := g \lor h$.
The same state receives a \code|fail| action guarded by $g \land h$.
Under this construction, the state rejects an execution that reaches a second fork, while allowing the first \code|fork| because the check uses the incoming value of $h$. 
A transaction that terminates without \code|fork|ing must still spawn a new transaction during the cycle when it exits, so at $q_{\mathrm{exit}}$, we add a \code|fork| action guarded by $\neg h$.

Second, \lang{} forbids forking in the first cycle; to enforce this we add \code|fail| guarded by $g$ whenever the entry state of the cycle-aligned automaton contains a \code|fork| guarded by $g$.
\input{fig/fork-check-example}

\section{Iterator Loops}
\label{app:iterator-loops}

\subsection{Driver Lowering}
\label{app:iterator-lowering}
\label{sec:iterator-lowering}
Lowering \code|for-in| and \code|repeat| loops requires special attention. A \code|repeat| loop is treated as syntactic sugar for a \code|for-in| loop over a client-provided stream of null elements; therefore, we only discuss \code|for-in| loops here.

For each invocation of a sequence parameter, we assume an abstract iterator $\ell$ with primitives $\ell.\operatorname{valid}()$, $\ell.\operatorname{data}()$, and $\ell.\operatorname{last}()$ that respectively indicate whether the next item from the iterator is available, the next item's value, and whether it is the final item.
These primitives are intrinsic to the compiler; Appendix~\ref{app:iterator-hardware} describes how hardware can supply them.
Each selected iteration consumes one element and loads the data register $s$ and validity register $\ell_{\mathrm{valid}}$, both initialized to zero.
These registers retain the sampled values until the next iteration, allowing the iterator to prepare subsequent items while the body executes without requiring its outputs to remain stable.

If a loop does not contain \code|is_last()|, we never read $\ell.\operatorname{last}()$; the loop executes its body while an element is available and exits otherwise.
If it contains \code|is_last()|, we specialize two body copies for nonfinal and final elements, replacing that expression with \code|false| and \code|true| respectively.
The nonfinal copy requires another element after its body completes, while the final copy exits after its entire body completes.
An empty sequence may exit at the initial entry, and no register for the final-element bit is needed.

If \code|iter_count()| is used, a counter register is initialized to 0 and incremented between iterations.

\begin{figure}[t]
\centering
\begingroup
\fontsize{9}{10.5}\selectfont
\input{fig/ast-to-ir-styles}
\begin{minipage}[t]{.32\linewidth}
\raggedright
  \code|for s in seq|
  \par (without \code|is_last()|):\par\smallskip
  \begin{tikzpicture}[x=1cm,y=1cm]
    \node[irvertex] (head) at (.35,0) {};
    \node[iraction, minimum width=1.9cm, minimum height=7mm,
          font=\fontsize{8}{9.5}\selectfont] (load) at (2.7,0) {
      $s:=\ell.\operatorname{data}()$\\
      $\ell_{\mathrm{valid}}:=\ell.\operatorname{valid}()$};
    \node[irtrue, minimum width=1.05cm] (body) at (2.7,-1.3) {$\mathcal B$};
    \node[irexits] (exit) at (.35,1.5) {};
    \draw[irnonstep] ([xshift=-3mm]head.west) -- (head.west);
    \draw[irtrueedge] (head) --
      node[irlabel,above=3.5mm,pos=.4,text=blue!65!black] {$\ell.\operatorname{valid}()$} (load);
    \draw[irnonstep] (load) -- (body);
    \draw[irfalseedge] (head.north) --
      node[irlabel,right=2pt,pos=.65,text=orange!75!black] {$\neg\ell.\operatorname{valid}()$} (exit.south);
    \draw[irnonstep] (body.west) -| (head.south);
    \draw[irnonstep] (exit) -- (3.9,1.5);
  \end{tikzpicture}
\end{minipage}\hfill
\begin{minipage}[t]{.665\linewidth}
\raggedright
  \code|for s in seq|\par
  (with \code|is_last()|):\par\smallskip
  \begin{tikzpicture}[x=1cm,y=1cm]
    \node[irvertex] (head) at (.4,0) {};
    \node[irnodelabel,anchor=north west] at (.05,-.48) {};
    \node[iraction, minimum width=1.9cm, minimum height=7mm,
          font=\fontsize{8}{9.5}\selectfont] (loadfinal) at (2.65,.75) {
      $s:=\ell.\operatorname{data}()$\\
      $\ell_{\mathrm{valid}}:=\ell.\operatorname{valid}()$};
    \node[iraction, minimum width=1.9cm, minimum height=7mm,
          font=\fontsize{8}{9.5}\selectfont] (loadbody) at (2.65,-.75) {
      $s:=\ell.\operatorname{data}()$\\
      $\ell_{\mathrm{valid}}:=\ell.\operatorname{valid}()$};
    \node[irtrue,font=\fontsize{8}{9.5}\selectfont] (finalbody) at (5.1,.75)
      {$\mathcal B[\operatorname{is\_last}()\mapsto1]$};
    \node[irtrue,font=\fontsize{8}{9.5}\selectfont] (body) at (5.1,-.75)
      {$\mathcal B[\operatorname{is\_last}()\mapsto0]$};
    \node[irexits] (exit) at (8.4,.75) {};
    \node[irvertex] (next) at (6.9,-.75) {};
    \node[iraction, minimum width=.95cm, minimum height=5mm] (fail) at (8.4,-.75)
      {$\mathrm{fail}$};
    \draw[irnonstep] ([xshift=-3mm]head.west) -- (head.west);
    \draw[irtrueedge] (head) -- (loadfinal);
    \node[irlabel,text=blue!65!black] at (1.25,1.45)
      {$\begin{gathered}\ell.\operatorname{valid}()\land\\\ell.\operatorname{last}()\end{gathered}$};
    \draw[irtrueedge] (head) -- (loadbody);
    \node[irlabel,text=blue!65!black] at (1.25,-1.45)
      {$\begin{gathered}\ell.\operatorname{valid}()\land\\\neg\ell.\operatorname{last}()\end{gathered}$};
    \draw[irfalseedge] (head.north) -- (.4,2.35) -- (8.4,2.35) -- (exit.north);
    \node[irlabel,above=1pt,text=orange!75!black] at (4.5,2.35) {$\neg\ell.\operatorname{valid}()$};
    \draw[irnonstep] (loadfinal) -- (finalbody);
    \draw[irnonstep] (loadbody) -- (body);
    \draw[irnonstep] (finalbody) -- (exit);
    \draw[irnonstep] (exit) -- (8.95,.75);
    \draw[irnonstep] (body) -- (next);
    \draw[irtrueedge] (next.south) -- (6.9,-1.65) -- (2.65,-1.65) --
      (loadbody.south);
    \node[irlabel,below=1pt,text=blue!65!black] at (4.7,-1.65)
      {$\ell.\operatorname{valid}()\land\neg\ell.\operatorname{last}()$};
    \draw[irtrueedge,preaction={draw=white,line width=2pt}]
      (next.north) -- (6.9,1.65) -- (2.65,1.65) -- (loadfinal.north);
    \node[irlabel,above=1pt,text=blue!65!black] at (4.7,1.65)
      {$\ell.\operatorname{valid}()\land\ell.\operatorname{last}()$};
    \draw[irfalseedge] (next) --
      node[irlabel,above=4mm,text=orange!75!black] {$\neg\ell.\operatorname{valid}()$} (fail);
  \end{tikzpicture}
\end{minipage}
\par\medskip
\endgroup
\caption{Syntax-directed lowering of \code|for-in| loops, without and with \code|is_last()|. Dashed arrows are nonstep edges. The body copies $\mathcal B[\operatorname{is\_last}()\mapsto0]$ and $\mathcal B[\operatorname{is\_last}()\mapsto1]$ specialize \texttt{is\_last()} to false and true, respectively.}
\label{fig:ast-to-ir-iterations}
\end{figure}

\subsection{Monitor Lowering}
\label{app:monitor-iterations}
\label{sec:monitor-iterations}
Recall that transition guards are parameter-free, so they can be evaluated from DUT ports, registers, and constants.
\code|for-in| loops have transitions based on $\ell.\operatorname{valid}()$, $\ell.\operatorname{data}()$, and $\ell.\operatorname{last}()$, which we do not have direct valuations of in the monitor.
Instead, monitoring a \code|for-in| loop requires reconstructing both the elements of the underlying list and the \emph{control flow} of the loop body, because the monitor has no supplied iterator.
To handle this, we determine which loop branch is active by checking its constraints against the current DUT sample, assuming at most one branch's constraints succeed in its first cycle.
If none succeeds, the monitor takes the empty-iterator branch if permitted by the length condition; otherwise, it rejects.

Recall the driver branches on the following cases:
\begin{enumerate}
  \item $\mathit{valid}\land\neg\mathit{last}$: a non-final item in the iterator
  \item $\mathit{valid}\land\mathit{last}$: the last item in the iterator
  \item $\neg\mathit{valid}$: empty iterator
\end{enumerate}

\begin{definition}[Iterator match]
At a state inspecting iterator $\ell$, let $g_\ell$ be the guard on all cases of entering the loop. Intuitively, this is the path condition $g_\ell$ for reaching the loop entry.
We parameterize by the possible values (0 or 1) of $\ell.\mathrm{valid}()$ and $\ell.\mathrm{last}()$, calling them $\mathit{valid}$ and $\mathit{last}$ respectively. 
Let $\mathcal C_\ell(\mathit{valid},\mathit{last})$ contain the guarded equality checks generated from the state's DUT-input assignments.
Sequences may have restrictions on their minimum length in \lang{}. Let $L_\ell(\mathit{valid},\mathit{last})$ be the length condition enforcing these, which we will defer detailed discussion of for now.

The conditions are given after the item-learning construction below.
Define the candidate's passing condition by
\[
  \operatorname{match}_\ell(\mathit{valid},\mathit{last})\triangleq
  \left(
    \bigwedge_{[g]\,\operatorname{assert\_eq}(e_1,e_2)\in\mathcal C_\ell(\mathit{valid},\mathit{last})}
      \bigl(\neg g\lor(e_1=e_2)\bigr)
  \right)
  \land\bigl(\neg g_\ell\lor L_\ell(\mathit{valid},\mathit{last})\bigr).
\]
\end{definition}

Now, we can replace uses of valid and last with instances of these matches. 
If the loop does not use \texttt{is\_last()} (and therefore not $\ell.\mathrm{last}()$), the plain loop has one candidate body, with passing condition
\[
  g_{\mathrm{body}}\triangleq\operatorname{match}_\ell(1,0).
\]
Replace $\ell.\mathrm{valid}()$ by $g_{\mathrm{body}}$ throughout the state's actions and transitions.

\paragraph{With \texttt{is\_last()}.}
Test the two specialized bodies using
\[
  g_{\mathrm{nonfinal}}\triangleq\operatorname{match}_\ell(1,0),
  \qquad
  g_{\mathrm{final}}\triangleq\operatorname{match}_\ell(1,1).
\]
Replace iterator functions by
\[
  \ell.\mathrm{valid}()\mapsto
    g_{\mathrm{nonfinal}}\lor g_{\mathrm{final}},
  \qquad
  \ell.\mathrm{last}()\mapsto g_{\mathrm{final}}.
\]
The nonfinal and final dispatch guards become $g_{\mathrm{nonfinal}}\land\neg g_{\mathrm{final}}$ and $g_{\mathrm{final}}$, respectively.

\paragraph{Learning each item.}
In $\mathbf{for}\ s\ \mathbf{in}\ seq\ \{B\}$, the loop variable $s$ is stored in a register also named $s$, as shown in Figure~\ref{fig:ast-to-ir-iterations}.
Each entry into the loop body introduces a fresh item, represented in the driver by $\ell.\mathrm{data}()$ and stored by $s:=\ell.\mathrm{data}()$.
Edge contraction substitutes $\ell.\mathrm{data}()$ for uses of $s$ following this assignment within the same cycle.
Uses across a step boundary instead refer to the register holding $s$, which contains the saved item.
A contracted state can contain both kinds of uses, for example when one iteration finishes and the next begins in the same cycle.

During conversion of each state, accumulate separate value and known-bit summaries for the fresh and saved items, initially
\[
\begin{aligned}
  (\mathit{Value}_{\mathrm{fresh}},\mathit{KnownBits}_{\mathrm{fresh}})&=(\{(1,0)\},\{(1,0)\}),\\
  (\mathit{Value}_{\mathrm{saved}},\mathit{KnownBits}_{\mathrm{saved}})&=(\{(1,v_s)\},\{(1,k_s)\}).
\end{aligned}
\]
The fresh item starts unknown, while the saved item inherits the persistent value and known-bit registers $v_s,k_s$.
Apply the observation rules with $\ell.\mathrm{data}()$ and $s$ as distinct parameter leaves: an equality with $\ell.\mathrm{data}()$ on one side checks and updates $(\mathit{Value}_{\mathrm{fresh}},\mathit{KnownBits}_{\mathrm{fresh}})$, while an equality with $s$ on one side checks and updates $(\mathit{Value}_{\mathrm{saved}},\mathit{KnownBits}_{\mathrm{saved}})$.

Let $g_{\mathrm{item}}$ be $g_{\mathrm{body}}$ in the plain case and $g_{\mathrm{nonfinal}}\lor g_{\mathrm{final}}$ in the specialized case.
Capture occurs under $g_{\mathrm{capture}}\triangleq g_\ell\land g_{\mathrm{item}}$.
After processing all observations, write the fresh item's value and known-bit summaries to $v_s,k_s$ if capture occurs.
Otherwise, write the saved item's summaries, including any bits learned in this state.
The two non-blocking register updates use the summaries below, where $\operatorname{coalesce}$ combines alternatives with identical expressions by disjoining their guards:
\[
\begin{aligned}
  V_{v_s}&\triangleq\operatorname{coalesce}
    \{(g_{\mathrm{capture}},\mathit{Value}_{\mathrm{fresh}}),
      (\neg g_{\mathrm{capture}},\mathit{Value}_{\mathrm{saved}})\},\\
  V_{k_s}&\triangleq\operatorname{coalesce}
    \{(g_{\mathrm{capture}},\mathit{KnownBits}_{\mathrm{fresh}}),
      (\neg g_{\mathrm{capture}},\mathit{KnownBits}_{\mathrm{saved}})\}.
\end{aligned}
\]

\paragraph{Length conditions.}
For a minimum sequence length $m$, maintain the number of required items still outstanding in a register $c_\ell$, initially $m$.
Table~\ref{tab:iterator-length-conditions} gives the length condition $L_\ell(\mathit{valid},\mathit{last})$.
\begin{table}[t]
\caption{Iterator length conditions, where $c_\ell$ counts the required items still outstanding.}
\label{tab:iterator-length-conditions}
\papertablestyle
\begin{tabular}{@{}ll@{}}
\toprule
Candidate case & Length condition $L_\ell(\mathit{valid},\mathit{last})$ \\
\midrule
Empty ($\mathit{valid}=0$) & $c_\ell=0$ \\
Plain-loop item ($\mathit{valid}=1$) & $1$ \\
Nonfinal item ($\mathit{valid}=1,\mathit{last}=0$) & $1$ \\
Final item ($\mathit{valid}=1,\mathit{last}=1$) & $c_\ell\le1$ \\
\bottomrule
\end{tabular}
\end{table}
Use the plain-loop row when \code|is_last()| is unused and the final two rows otherwise.
The counter is tested before capture, so a final item is allowed when the minimum is already satisfied ($c_\ell=0$) or this item satisfies the last outstanding requirement ($c_\ell=1$).
The length check is guarded by $g_\ell$ and gates \code|fork| actions and transitions, including those reaching $q_{\mathrm{exit}}$.
Decrement on capture with saturation at zero:
\[
  V_{c_\ell}\triangleq
  \{(g_{\mathrm{capture}}\land(c_\ell>0),c_\ell-1),
    (\neg g_{\mathrm{capture}}\lor(c_\ell=0),c_\ell)\}.
\]
A sequence with minimum length zero needs no remaining-item register.

\subsection{Implementing Iterators in Hardware}
\label{app:iterator-hardware}
The hardware driver requires a source of sequence items that obeys the iterator contract. A FIFO is one implementation: its outputs \texttt{t\_valid}, \texttt{t\_last}, and \texttt{t\_data} supply $\ell.\operatorname{valid}()$, $\ell.\operatorname{last}()$, and $\ell.\operatorname{data}()$, respectively. Whenever the loop selects an iteration, \texttt{t\_valid} indicates that an item is available, \texttt{t\_data} gives that item, and \texttt{t\_last} is high exactly for the final item. The source advances once when the driver consumes an item. The driver latches the selected item, so the source need not hold it throughout the loop body. An empty sequence presents \texttt{t\_valid} low at the initial loop entry.

A \code|repeat| loop needs only a stream of $n$ dummy items. A counter can provide \texttt{t\_valid} while fewer than $n$ iterations have been consumed and raise \texttt{t\_last} on the $n$th item; for $n=0$, it presents \texttt{t\_valid} low immediately. The value on \texttt{t\_data} is arbitrary because the loop body never reads it. Thus, this source needs no FIFO storage.

\section{Parallel Merge}
\label{app:parallel-merge}
\label{sec:parallel-merge-rules}
There are two key properties we would like to preserve of \lang{}'s semantics.
First, if one transaction has a concrete assignment while another assigns \code|X|, the concrete assignment always takes precedence. The parallel assignment merge provides us these semantics.

\begin{definition}[Parallel assignment merge]
Let $V_p$ and $V'_p$ be value summaries for the same symbol in two states executed by concurrent transactions, and let $\mathcal E_{V,V'}$ be the set of concrete expressions appearing in either summary.
Let $G_e$ be the guard of expression $e$ in $\operatorname{coalesce}(V_p\cup V'_p)$, or $0$ if $e$ is absent.
The conflict guard $\kappa(V_p,V'_p)$ detects simultaneous selection of distinct concrete expressions:
\[
  \kappa(V_p,V'_p)\triangleq
    \bigvee_{\substack{e,e'\in\mathcal E_{V,V'},\:\: e\ne e'}}
      (G_e\land G_{e'}).
\]
Because each input summary has disjoint guards, such a conflict can only arise between the two summaries.
Let $C\triangleq\bigvee_{e\in\mathcal E_{V,V'}}G_e$.
The \emph{parallel assignment merge} is
\[
  V_p\mathbin{\bowtie}V'_p
  \triangleq
  \{(G_e\land\neg\kappa(V_p,V'_p),e)\mid e\in\mathcal E_{V,V'},\ e\ne\texttt{X}\}
  \cup
  \{((G_{\texttt{X}}\land\neg C)\lor\kappa(V_p,V'_p),\texttt{X})\}.
\]
Note that the resulting guards remain pairwise disjoint.
\end{definition}

Second, we conservatively treat simultaneous assignments of distinct concrete expressions to the same port as data races, even when those expressions happen to evaluate to the same value.
To do this, when we merge assignments via parallel merge, we also emit a \code|fail| action under $\kappa(V_p,V'_p)$. All fail actions are disjuncted, and assertions are simply appended.

\paragraph{Monitors.}
The same merge rules apply to monitor actions; because instance-qualified register assignments have disjoint targets, their parallel merge reduces to a union.
For each newly launched transaction, substitute its initial register values before constructing learning expressions and checks, so values learned by an earlier transaction cannot constrain it.
Record the same permutation and initialization slots as for the driver, but initialize the registers used to learn arguments rather than supplying arguments externally.

\section{Algorithm for Constructing the Configuration Automaton}
\label{app:meta-construction}
\label{sec:control-dependent-overlap}

The following worklist algorithm constructs the \meta{} described in \S\ref{sec:meta-automata}.
The worklist contains configurations whose successors have not yet been explored.
For each protocol $P$, write $q_i^P$ for state $i$ of its clock-aligned single-transaction automaton, and $f_i^P$ for the guard of that state's \code|fork| action, or $0$ when the action is absent.
Initialize $\mathcal C$ and the worklist to $\{C_0\}$, where $C_0$ is the frontier choice among the protocols' entry states, each with instance number $1$.
Initialize $\leadsto$ to the empty relation.
For each configuration removed from the worklist, construct its successors as follows:
\begin{enumerate}
  \item \emph{Enumerate possibilities and prune.}
  If the frontier is a choice, consider each protocol $P$ in that choice separately and record $P$ on the resulting transitions.
  Otherwise, record $\emptyset$.
  For each such case, enumerate the combinations obtained by taking one outgoing transition of the single-transaction automaton for each transaction currently executing.
  Let $g_0$ be the conjunction of the selected transition guards, with parameter and register references qualified by protocol and instance number.
  If the frontier reaches state $q_j^P$, split the combination into the cases $g=g_0\land f_j^P$ and $g=g_0\land\neg f_j^P$.
  The selected transition guards need not determine whether the fork guard holds.
  Without a frontier, use $g=g_0$.
  Discard combinations whose condition we can show to be unsatisfiable.
  Use $g$ as the guard of the resulting \meta{} transition.

  \item \emph{Process forks and retire transactions.}
  For each remaining combination, apply both rules:
  \begin{enumerate}
    \item If the frontier forks, remove it from the frontier and add a frontier choice as its child transaction, with one alternative at the entry state of each protocol and a fresh instance number for that protocol.
    \item Retire each transaction that reaches its exit state.
    Keep every other transaction at its reached state, retaining its overline if it is still the frontier.
  \end{enumerate}

  \item \emph{Form the successor configuration.}
  Conjoin the transactions produced above to form a candidate successor configuration.
  If all transactions have retired, finish this case without adding a configuration.

  \item \emph{Reuse an equivalent configuration.}
  Search $\mathcal C$ for a configuration $C'$ that matches the candidate by equivalence up to instance numbers.
  If a match exists, use it and take the mapping of surviving transactions' instance numbers from the match.
  Otherwise, use the candidate as $C'$ and add it to $\mathcal C$ and the worklist.
  Complete the mapping of surviving transactions to a permutation of instance slots within each protocol, using retired or unused slots for the remaining mappings.
  The slot pool must be large enough for the source and successor configurations; mappings involving only unused slots may be omitted.
  Add $C\overset{g,P}{\leadsto}C'$, using $P=\emptyset$ when no protocol is selected, and record the permutation on that transition.
  Separately record each slot $j$ of protocol $P$ assigned to a new child alternative in $C'$ for initialization after the permutation.
\end{enumerate}

By Proposition~\ref{prop:meta-construction-termination}, there are finitely many configurations up to instance numbers under the bounded post-fork phase condition.
The worklist adds a configuration only when none already in $\mathcal C$ matches it up to instance numbers.
Each configuration has finitely many successor cases, so the algorithm terminates and produces a finite \meta{}.

\newpage

\section{Reconstructor performance comparison}

\label{sec:fast-bi-ethmac}
\input{fig/fast-bi-timing-ethmac}

%% file: fig/fork-check-example.tex
\begin{figure}[t]
\centering
\begingroup
\tikzset{
  forkstate/.style={draw=black!70, fill=black!3, rounded corners=1pt,
    align=left, inner sep=3pt, font=\fontsize{8}{9.5}\selectfont},
  forkedge/.style={-{Latex[length=1.3mm]}, line width=.65pt},
}
\begin{tikzpicture}
  \node[forkstate,line width=1.2pt] (a0) at (0,0) {
    $\mathrm{D.a}:=a;\quad \mathrm{D.b}:=b$\\
    $\mathrm{D.op}:=0;\quad \mathrm{D.go}:=0$\\
    $v_a:=a;\quad v_b:=b$\\
    $v_{\mathrm{op}}:=0;\quad v_{\mathrm{go}}:=0$\\
    $x_a:=0;\quad x_b:=0$\\
    $x_{\mathrm{op}}:=0;\quad x_{\mathrm{go}}:=0$};
  \node[forkstate, right=6mm of a0] (a1) {
    $\mathrm{D.a}:=\texttt{X};\quad \mathrm{D.b}:=\texttt{X}$\\
    $\mathrm{D.op}:=\texttt{X};\quad \mathrm{D.go}:=\texttt{X}$\\
    $x_a:=1;\quad x_b:=1$\\
    $x_{\mathrm{op}}:=1;\quad x_{\mathrm{go}}:=1$\\
    $\operatorname{assert\_eq}(\mathrm{D.s},s)$\\
    $\mathrm{fork};\quad {\color{blue!65!black}[h]\,\mathrm{fail};\quad h:=1}$};
  \node[forkstate, right=6mm of a1] (a2) {
    $\mathrm{D.a}:=\{(x_a,\texttt{X}),(\neg x_a,v_a)\}$\\
    $\mathrm{D.b}:=\{(x_b,\texttt{X}),(\neg x_b,v_b)\}$\\
    $\mathrm{D.op}:=\{(x_{\mathrm{op}},\texttt{X}),(\neg x_{\mathrm{op}},v_{\mathrm{op}})\}$\\
    $\mathrm{D.go}:=\{(x_{\mathrm{go}},\texttt{X}),(\neg x_{\mathrm{go}},v_{\mathrm{go}})\}$\\
    ${\color{blue!65!black}[\neg h]\,\mathrm{fork}}$};
  \draw[forkedge] (a0) -- (a1);
  \draw[forkedge] (a1) -- (a2);
\end{tikzpicture}
\endgroup
\caption{The \code|add| driver instrumented with assignment and fork histories.}
\label{fig:add-d1-fork-check}
\end{figure}

%% file: fig/fast-bi-timing-ethmac.tex
\begin{figure}[htp]
\centering
\includegraphics[width=0.8\linewidth]{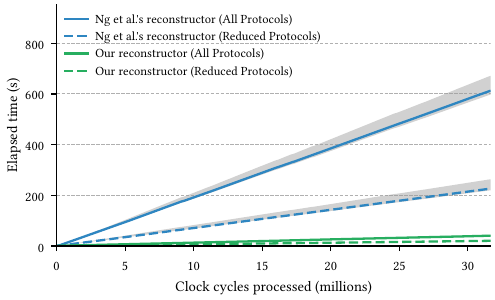}
\caption{Wall-clock time of Ng et al.'s \cite{ng_et_al_2026} and our reconstructors on an OpenCores Ethernet Wishbone waveform containing over 31 million cycles (lower is better). 
We plot the mean wall-clock time over 10 runs, with the minimum and maximum values indicated by gray bands (note that the gray bands are too tight to be visible for our reconstructor).}
\label{fig:fast-bi-timing-ethmac}
\end{figure}